\documentclass[a4paper,11pt]{article}
\pdfoutput=1
\usepackage{jcappub}

\usepackage{amsmath}
\usepackage{amsthm}
\usepackage{amssymb}
\usepackage{amsfonts}
\usepackage{mathtools}

 \usepackage[english]{babel}
\usepackage{lineno}
\usepackage{lscape}
\usepackage{soul}
\usepackage{graphicx}
\usepackage{bm}
\usepackage{epstopdf}
\usepackage{url}
\usepackage{booktabs}
\usepackage{mdframed}
\usepackage[colorinlistoftodos]{todonotes}
\usepackage{booktabs}
\usepackage{graphicx}
\usepackage{booktabs}
\usepackage[super]{nth}
\usepackage[math]{cellspace}

\usepackage{fontawesome}
\usepackage{microtype}
\usepackage{xspace}
\usepackage[capitalise]{cleveref}
\crefname{section}{Sec.}{Secs.} \Crefname{section}{Section}{Sections.}
\usepackage{verbatim}
\usepackage{enumitem}
\usepackage{siunitx}
\usepackage{hyperref}
\usepackage{caption}
\usepackage{booktabs,tabularx}
\usepackage{subcaption}
\DeclareSIUnit\parsec{pc}
\DeclareSIUnit\year{yr}
\DeclareSIUnit\solarmass{\textit{M}_\odot}

\newcommand{\IFCA}{%
Instituto de F\'isica de Cantabria (IFCA), University of Cantabria (UC)-CSIC,\\
 Avenida de los Castros, s/n E-39005 Santander, Spain}

\newcommand{\Unipi}{%
Dipartimento di Fisica E. Fermi, Universit\`a di Pisa, Largo B. Pontecorvo 3, I-56127 Pisa, Italy }

\newcommand{\INFN}{%
INFN, Sezione di Pisa, Largo Bruno Pontecorvo 3, I-56127 Pisa, Italy}

\preprint{}

\begin{document}

\title{The cosmic history of Primordial Black Hole accretion and its uncertainties}

\author[a]{Pratibha Jangra,}
\author[b,c]{Daniele Gaggero,}
\author[a]{Bradley J. Kavanagh,}
\author[a]{J. M. Diego}

\affiliation[a]{\IFCA}
\affiliation[b]{\Unipi}
\affiliation[c]{\INFN}

\emailAdd{pratibha@ifca.unican.es}
\emailAdd{dgaggero@pi.infn.it}
\emailAdd{kavanagh@ifca.unican.es}
\emailAdd{jdiego@ifca.unican.es}


\abstract{Primordial Black Holes (PBHs) have not been experimentally detected so far, but their existence would provide important insights about the early Universe and serve as one of the possible candidates of dark matter (DM). In this work, we explore the accretion of radiation and matter by PBHs, with relevance for the growth of PBH seeds to form early Supermassive Black Holes; the emission from accreting PBHs; and constraints from gravitational wave observations, among others. We study the growth of PBH masses in the early Universe due to the accretion of radiation, highlighting uncertainties which arise from estimates of the PBH formation time. For baryonic accretion, we review the traditional Bondi-Hoyle-Lyttleton (BHL) and its refined version known as the Park-Ricotti (PR) model, which also includes radiative feedback. We find that in the BHL model, PBHs heavier than $\sim 100 \,\mathrm{M_{\odot}}$ can grow in mass by several orders of magnitude by $z \lesssim 10$, though only when surrounded by DM halos and only when the accretion efficiency is large. By contrast, the inclusion of radiation feedback in the PR model can drastically suppress the baryonic accretion rate of PBHs, leading to a negligible change in PBH mass over cosmic time. Furthermore our calculations show that the accretion rate depends sensitively on the modelling of various parameters such as the speed of sound in the baryonic gas and the velocity of PBHs. These findings highlight the uncertainties associated with accretion onto PBHs, and we find that a large increase in the PBH mass due to accretion is by no means guaranteed.}

\maketitle
\section{Introduction}
\label{sec:Introduction}
With the detection of gravitational waves by the LIGO-Virgo Collaboration~\cite{LIGOScientific:2016aoc}, the plausible existence of Primordial Black Holes (PBHs) has earned them major attention. As evident from the terminology, PBHs have a primordial origin, forming from the collapse of large over-densities in the early Universe~\cite{Zeldovich:1967lct,Hawking:1971ei,Green:1999xm,Green:2020jor,Villanueva-Domingo:2021spv,Carr:2021bzv,Carr:2023tpt}. Predictions have been made about the merger rates of PBHs in binaries~\cite{Bird:2016dcv,Clesse:2016vqa,Sasaki:2016jop}, which can potentially be probed by the improving sensitivity of the LVK Collaboration and future detectors such as Einstein Telescope~\cite{Branchesi:2023mws}. A key goal of these studies is to constrain  the abundance $f_\mathrm{pbh}$ of PBHs as a non-particle fraction of dark matter~\cite{Sasaki:2016jop,Raidal:2018bbj,Kavanagh:2018ggo,DeLuca:2020qqa,Hall:2020daa,Hutsi:2020sol,Wong:2020yig,DeLuca:2020jug,Bhagwat:2020bzh,Franciolini:2021tla,DeLuca:2021wjr,Deng:2021ezy,Chen:2021nxo,Chen:2022fda,Liu:2022iuf,Postnov:2023ntu} and to fully comprehend their evolution in the Universe, post-formation.

Since accretion around PBHs can alter their mass as well as their spin~\cite{DeLuca:2020bjf}, so it is an important factor affecting the overall evolution of PBHs. Accretion around PBHs can also lead to the emission of radiation such as X-rays and radio-waves. So, depending on their abundance, early accretion around PBHs can leave significant impacts on the recombination history of the Universe. This might subsequently impact the properties of the Cosmic Microwave Background (CMB)~\cite{1981MNRAS.194..639C,Ricotti:2007au,Serpico:2024cdz}. Recent studies have also suggested that PBHs in the mass range $\mathcal{O}(1) - 10^{7} \, \mathrm{M_{\odot}}$ could emit detectable X-ray and radio signals if they accrete interstellar gas~\cite{Gaggero:2016dpq,Inoue:2017csr,Manshanden:2018tze,Serpico:2020ehh,Lu:2020bmd}. Instruments like the Very Large Array (VLA) and the Chandra X-ray Observatory can potentially probe these signals~\cite{Gaggero:2016dpq,Manshanden:2018tze}. Furthermore, studies of wide binaries and open clusters limits the abundance of PBHs heavier than $10 \, \mathrm{M_{\odot}}$~\cite{Hektor:2018qqw,Tashiro:2021kcg}, while the observations of the 21cm signal claimed from the EDGES experiment and CMB anisotropies constrain more massive PBHs~\cite{Hektor:2018qqw,Tashiro:2021kcg}. Additionally, accretion around PBHs of mass $\gtrsim 10^{3}\, \mathrm{M_{\odot}}$ has been proposed to explain the early presence of supermassive black holes (SMBHs)~\cite{1984MNRAS.206..801C,Bean:2002kx,2012Sci...337..544V,Inayoshi:2019fun} which are being observed in large numbers by the James Webb Space Telescope (JWST)~\cite{Cappelluti:2021usg,Natarajan:2023rxq,2024ApJ...966..176Y}. Moreover, the waveforms of GWs produced by merging PBH binaries may be significantly impacted if accretion disks are present~\cite{Ali-Haimoud:2017rtz,Cole:2022yzw,Becker:2022wlo}. So, the detailed study of accretion around PBHs is important to fully explore these observations and predictions.

Many efforts have been made to describe the phenomenon of accretion taking place around PBHs. For example, Ref.~\cite{Nayak:2009wk} showed that the spherical accretion of radiation around very light PBHs can halt their evaporation process, leading more PBHs to survive until today. Reference~\cite{Nayak:2011sk} concluded that the accretion of radiation before matter-radiation equality (MRE) can increase the mass of PBHs up to $40\%$, while the accretion of matter during matter-domination (MD) can increase the mass of PBHs by up to $90 \%$. Similar results were obtained for the accretion of radiation by rotating PBHs~\cite{Mahapatra:2013bpa}. However, in both Refs.~\cite{Nayak:2009wk,Nayak:2011sk}, the redshift dependence of the accretion efficiency $\lambda$ and the contribution of the redshift of PBH formation were not comprehensively accounted for, in order to accurately estimate the change in PBH mass via accretion.

For baryonic accretion, Ricotti et al.~\cite{Ricotti:2007au} (hereafter \textit{ROM07}) employed the traditional Bondi-Hoyle-Lyttleton (BHL) model~\cite{hoylelyttleton1939,Bondi:1944jm} for PBHs, using a model for the accretion efficiency developed in Ref.~\cite{Ricotti:2007jk}. They predicted that PBHs gain mass via spherical accretion of baryons which is further enhanced by the presence of particle dark matter halos around isolated PBHs~\cite{1984ApJ...281....1F,1985ApJS...58...39B,Mack:2006gz,Adamek:2019gns,Boudaud:2021irr,Jangra:2023mqp}. Using the formalism of ROM07, Ref.~\cite{DeLuca:2020fpg} showed that for massive PBHs with DM halos, accretion can increase their mass by many orders of magnitude, at redshift $z \leq 15$, with similar findings in other works~\cite{Rice:2017avg}. Reference~\cite{DeLuca:2020fpg} also predicted that accretion by PBHs with mass $\geq 0.1 \,\mathrm{M_{\odot}}$ can weaken the existing constraints on their abundance signifying that PBHs could constitute a large fraction of dark matter. However, the BHL framework applied in these works does not take into account the effects of radiative feedback. Furthermore, as we will explore later in this work, this formalism is subject to substantial uncertainties coming from assumptions about the sound speed and PBH velocities. Serpico et al.~(\cite{Serpico:2020ehh}, hereafter \textit{SPIK20}) presented an updated BHL formalism for the accretion rates of PBHs post-recombination, emphasizing the possibilities of disk formation around PBHs. They concluded that in comparison to spherical accretion, the formation of accretion disks around PBHs leads to higher luminosities, which may enhance the effects of feedback and therefore alter the rate of accretion. However, the effects of the potential infall of the PBHs into virialized halos during the non-linear regime were not taken into account in Ref.~\cite{Serpico:2020ehh}. Hence, in spite of significant advancements, many uncertainties still exist in the formalism of accretion dynamics for PBHs and its subsequent consequences for their evolution.

In this work, we present a refined theoretical framework for PBH accretion, exploring uncertainties arising from the choice of accretion model; the presence of particle DM halos around PBHs; and the PBH and gas velocities. One aspect of this involves an updated formulation of the BHL model for radiation as well as baryonic accretion onto PBHs with and without DM halos, mentioned in Refs.~\cite{Nayak:2009wk,Nayak:2011sk,Ricotti:2007jk,Ricotti:2007au,DeLuca:2020fpg,Serpico:2020ehh}. We analyze radiation accretion from deep radiation domination (RD) up to MRE, self-consistently accounting for the time of formation of the PBHs. We subsequently study baryonic accretion down to low redshifts, using the ROM07 model for the accretion efficiency, speed of sound and the velocity of PBHs. In contrast to ROM07, we use an alternative formulation by directly including the influence of DM halos to generalise the Bondi radius - the region of effective gravitational capture - around PBHs. 
We then present a novel analysis of cosmic accretion for PBHs, based on the Park-Ricotti (PR) accretion model outlined in Refs.~\cite{Park:2010yh,Park:2011rf,Park:2012cr,Sugimura:2020rdw,Scarcella:2020ssk}. This takes into account the effects of radiative feedback resulting in a higher speed of sound localized within the region ionized by the trapped radiation. A similar approach is presented in Ref.~\cite{Agius:2024ecw} which explored the impact of PBH accretion (in both the BHL and PR models) on the CMB, and subsequent constraints on the PBH abundance. Here, we incorporate the redshift dependence of the accretion efficiency $\lambda$ (more specifically, the effects of gas viscosity and Hubble expansion influencing the efficiency of accretion) and changes in PBH velocity due to late-time structure formation. With this, we study the evolution of PBH masses over cosmic time in the PR model, presenting a comprehensive analysis and comparison with the previously studied BHL model.

Additionally, we also examine how the accretion rates of PBHs depend on various accretion parameters, particularly the choice of sound speeds and PBH velocities over cosmic time. We use two sets of these `velocity profiles' from ROM07 and SPIK20, and analyze their impact on cosmic accretion. A similar approach is explored in Ref.~\cite{Banerjee:2024qzl}, which discusses the accretion rates of PBHs for different cosmological models, in the context of theories of modified gravity. Here, we aim instead to highlight the sensitivity of accretion to modeling the cosmological dynamics within the context of $\Lambda$CDM.

The paper is organized as follows: First, in Sec.~\ref{sec:Accretion of radiation around PBHs}, using the BHL model, we estimate the changes in the mass of PBHs by the accretion of radiation in RD. Then, in Sec.~\ref{sec:Bondi-Hoyle-Lyttleton (BHL) Accretion model for baryons}, we study baryonic accretion around PBHs with and without the presence of DM halos in the BHL model. After that, in Sec.~\ref{sec:Park Ricotti (PR) Accretion model}, we refine our analysis by applying the framework of the PR model, which also includes the effects of radiative feedback. In Sec.~\ref{subsec:Accreted masses of PBHs and their Velocity dependence}, we present our results for the final PBH masses, comparing the results of the BHL and PR models. We also summarise there the impact of uncertainties in the cosmic velocity profiles, leaving the detailed analysis for Appendix~\ref{sec:Sensitivity of the accretion rate on the choice of velocity profiles}.
Finally, in Sec.~\ref{sec:Conclusion}, we conclude our work discussing the scope of further improvements. Through the detailed analysis presented in this paper, we aim to narrow down the existing gap between the theoretical uncertainties and the observational constraints, advancing our understanding of PBHs and their role in the Universe.

\section{Accretion of radiation around PBHs}
\label{sec:Accretion of radiation around PBHs}
Accretion is the process of the gravitational capture of the surrounding material by nearby compact objects. It was first studied by Hoyle and Lyttleton~\cite{hoylelyttleton1939} in 1939, and was later updated by Bondi and Hoyle~\cite{Bondi:1944jm,10.1093/mnras/112.2.195}, encapsulated as the Bondi-Hoyle-Lyttleton (BHL) model. 
This model assumes that the surrounding material is uniformly distributed out to infinity and neglects self-gravity. The BHL model asserts that particles are accreted from within the Bondi radius~\cite{hoylelyttleton1939,Bondi:1944jm}, the radius at which the sound-speed in the gas becomes equal to the escape velocity from the compact object. This represents the boundary between regions of sub-sonic and super-sonic infall. These key assumptions in the BHL model lay the foundation to robustly describe the various processes where accretion plays an important role, ranging from star formation to the evolution of black holes~\cite{Edgar:2004mk}.

Within the Bondi-Hoyle-Lyttleton (BHL) model, the rate of accretion for an isolated PBH of mass $M$ is given by~\cite{hoylelyttleton1939,Bondi:1944jm}:
\begin{equation} \label{eq:Mdot} 
    \dot M  = 4  \pi \lambda \, \rho \,v_\mathrm{eff} \, r_\mathrm{B}^{2}\,,
\end{equation}
where $r_\mathrm{B}$ is the Bondi radius defined as:
\begin{equation} \label{eq:bondiradius}
    r_\mathrm{B} = \frac{G M}{v_\mathrm{eff}^{2}}\,,
\end{equation}
with effective velocity 
\begin{equation} \label{eq:veff}
   v_\mathrm{eff} = \sqrt{v_\mathrm{pbh}^{2} + c_{s}^{2}}\,.
\end{equation}
Here, $v_\mathrm{pbh}$ is the proper velocity of the PBHs and $c_s$ is the speed of sound in the surrounding medium of density $\rho$. Additionally, the parameter $\lambda$ is the accretion efficiency. Historically, $\lambda$ was introduced to address discrepancies between the theoretically predicted accretion rates of compact objects with the observed data~\cite{hoylelyttleton1939,Bondi:1944jm}. The maximum allowed value of $\lambda $ is unity signifying the ideal scenarios of accretion while depending on the modeling conditions, its typical values lie in the range $0.001 - 0.1$~\cite{Perna:2003ck,Ricotti:2007au,Ali-Haimoud:2016mbv,10.1093/mnras/sts688,Agius:2024ecw}.
In this section, we focus on the accretion of radiation in the early Universe. For simplicity, in our study, we choose $\lambda = 0.1$ for the accretion of radiation around PBHs up to MRE.

Since the formation of PBHs takes place just after the end of inflation, so it allows them to begin accreting radiation right after coming into existence. 
The initial mass of the PBHs at the time $t_i$ of their formation can be expressed as~\cite{Cole:2017gle}:
\begin{equation} \label{eq:Minitial}
    M_i = \gamma M_\mathrm{H} =  \frac{\gamma c^{3}}{2 G  H(t_{i})} \,.
\end{equation}
Here, $M_\mathrm{H}$ is the horizon-mass,  $\gamma$ is the ratio of the horizon and PBH masses~\cite{Green:1999xm,Green:2020jor,Villanueva-Domingo:2021spv} and $H (t_i) = 1/2t_i$ is Hubble parameter in RD at the time $t_i$ of PBH formation. The value of $\gamma$ depends on the shape of the the density fluctuations leading to the formation of the PBHs. For the most commonly studied perturbations, such as Gaussian and Mexican Hat over-densities in Fourier-space, it is usually considered that $\gamma\approx 0.37$~\cite{Green:1999xm}. Other studies also suggest somewhat lower values, such as $\gamma \simeq 0.2$~\cite{1975ApJ...201....1C}. In a universe containing both matter and radiation, the cosmic time can be written as~\cite{Jangra:2023mqp}:
\begin{equation}  \label{eq:time}
     t = \left(\frac{3}{4\pi G \rho_\mathrm{eq}}\right)^{1/2}\, \left[\frac{2}{3}\left(s-2\right)\left(s+1\right)^{1/2} + \frac{4}{3}\right]\,,
 \end{equation}
with $s = a/a_\mathrm{eq}$ being the scale-factor with respect to MRE. For deep RD, if we Taylor expand the above equation about $s = 0$ up to second order, we obtain:
 \begin{equation}  \label{eq:t_RD}
      t_\mathrm{RD}  =   \left(\frac{3}{4\pi G\rho_\mathrm{eq}}\right)^{1/2} \frac{s^{2}}{2}\,.
 \end{equation}
Then, substituting $a = 1/(1+z)$, the redshift of PBH formation can be expressed as:
\begin{equation} \label{eq:zi}
   z_i = \frac{1}{\sqrt{2 t_i}} \, \left(\frac{3}{4 \pi G\rho_\mathrm{eq}}\right)^{1/4}\,    \left( 1 + z_\mathrm{eq}\right) - 1\: \,.
\end{equation}
In order to determine $\mathrm{d}M/\mathrm{d}z = \dot{M} (\mathrm{d}t/\mathrm{d}z)$, we also use Eq.~\eqref{eq:time} to evaluate:
\begin{equation} \label{eq:dtdz}
    \frac{\mathrm{d}t}{\mathrm{d}z}=\sqrt{\frac{3}{8\pi G\rho_{\mathrm{c},0}}}\left(\frac{1}{\sqrt{ \Omega_{\mathrm{r},0}\, (1 + z)^{6} + \Omega_{\mathrm{m},0}\, (1 + z)^{5}}}\right)\,,
\end{equation}
where $\Omega_{\mathrm{r},0} = 9.4\times10^{-5},\:\Omega_{\mathrm{m},0} = 0.32,\:\Omega_{\Lambda,0} = 0.68, \: \rho_{\mathrm{c},0} = 1.9\times10^{-29}h^{2}\mathrm{g\,cm^{-3}}$ and $h = 0.67$. 
Then, using Eqs.~\eqref{eq:Mdot}~and~\eqref{eq:dtdz}, and setting $\rho \equiv \rho_{r} = \rho_{r,0} (1+z)^{4}$, we obtain:
\begin{equation} \label{eq:dMdz}
    \frac{\mathrm{d} M}{\mathrm{d} z} = - \left(\frac{4\pi \lambda G^{2}}{c_{s}^{3}}\right) \left(\frac{3}{8 \pi G \rho_{\mathrm{c},0}}\right)^{1/2} \,  \, \frac{M^{2}\, \rho_{r}}{\sqrt{ \Omega_{\mathrm{r},0}\, (1 + z)^{6} + \Omega_{\mathrm{m},0}\, (1 + z)^{5}}}\,.
\end{equation}
We fix the sound speed to $c_s = c/\sqrt{3}$ in RD~\cite{Rice:2017avg}.

\begin{figure}[tb!]
\centering
\includegraphics[width = 0.6\textwidth]{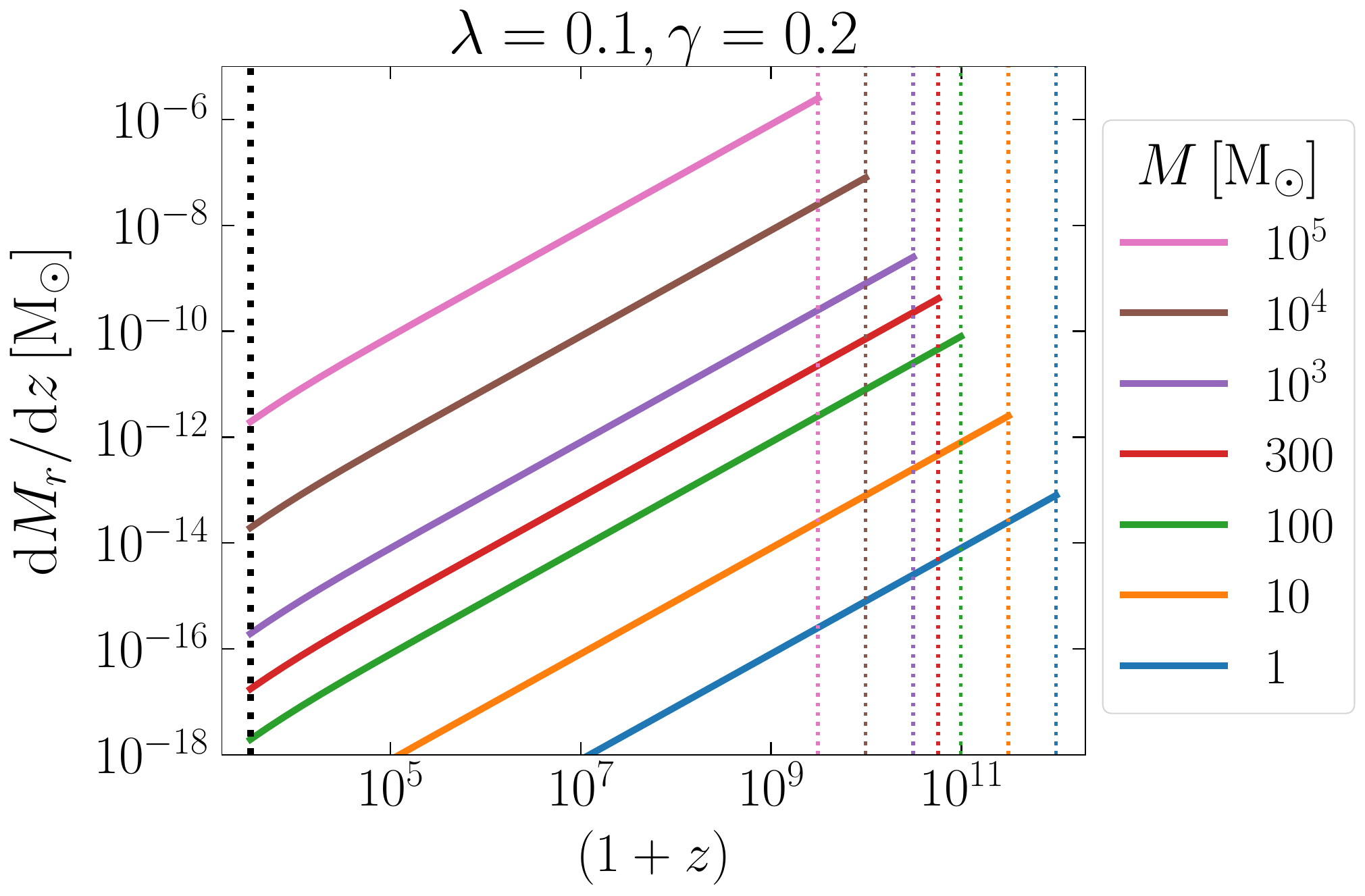}
\caption{Variation of the accretion rate of radiation around isolated PBHs, as a function of redshift $(1+z)$. The colored vertical dotted lines show the redshift of formation for PBHs with initial masses mentioned in the legend of the plot. The vertical black dotted line indicates the redshift of matter-radiation equality (MRE) i.e.\ $z_\mathrm{eq}\approx 3400$.}
\label{fig:racc1}
\end{figure}

The variation of the accretion rate of radiation around isolated PBHs is shown in Figure~\ref{fig:racc1}.
This figure shows that for accretion efficiency $\lambda = 0.1$ and $\gamma = 0.2$, the accretion rates of radiation around isolated static PBHs are high at higher redshift and then they decrease rapidly towards lower redshift. This is due to the fact that the density of radiation dilutes very quickly as $\propto (1+z)^{4}$. By the time of MRE, the accretion of radiation becomes negligible.

In order to evaluate the final PBH mass, we can rewrite Eq.~\eqref{eq:dMdz} as: 
\begin{equation}  \label{eq:3}
   \int^{M_{f,r}}_{M_i}  \frac{\mathrm{d} M}{M^{2}}  = - \left(\frac{4\pi \lambda G^{2}}{c_{s}^{3}}\right) \left(\frac{3}{8 \pi G \rho_{\mathrm{c},0}}\right)^{1/2} \rho_{r,0} \int^{z}_{z_i} \frac{(1 + z)^{4} \,\mathrm{d} z}{\sqrt{ \Omega_{\mathrm{r},0}\, (1 + z)^{6} + \Omega_{\mathrm{m},0}\, (1 + z)^{5}}} \,.
 \end{equation}
If we define a parameter $ \Bar{z} = \Omega_{\mathrm{r},0}(1 + z)/\Omega_{\mathrm{m},0}$, then the above equation can be expressed as:
\begin{equation}  \label{eq:Mfradintegral}
   \int^{M_{f,r}}_{M_i}  \frac{\mathrm{d} M}{M^{2}}  = - \;  \,   \sqrt{\frac{\Omega_{\mathrm{r},0}^{5}}{\Omega_{\mathrm{m},0}^{6}}}\,\left(\frac{\Omega_{\mathrm{m},0}}{\Omega_{\mathrm{r},0}}\right)^{5} \, \mathcal{C}\,\rho_{r,0} \int^{\Bar{z}}_{\Bar{z}_i} \frac{\Bar{z}^{4}}{\sqrt{  \Bar{z}^{6} +  \Bar{z}^{5}}}\,\mathrm{d} \Bar{z}\,,
 \end{equation}
with $\mathcal{C} = \left(\frac{4\pi \lambda G^{2}}{c_{s}^{3}}\right) \left(\frac{3}{8 \pi G \rho_{\mathrm{c},0}}\right)^{1/2}$ having dimensions of $\mathrm{[mass]}^{-2} \, \mathrm{[length]}^{3}$.
The integral over $\bar{z}$ can be evaluated analytically as:
 \begin{equation}  \label{eq:integral}
    \int_0^{\bar{z}} \frac{\Bar{z}^{4}}{\sqrt{ \Bar{z}^{6} + \Bar{z}^{5}}}\,\mathrm{d} \Bar{z} =  \frac{3\,\sqrt{(1 + \Bar{z})\Bar{z}^5}\, \mathrm{sinh}^{-1}(\Bar{z}) + (2\,\Bar{z}^{2} - \Bar{z} - 3)\, \Bar{z}^{3}}{4 \sqrt{(1 + \Bar{z})\Bar{z}^5}} \equiv \mathcal{I}(\Bar{z})\,.
 \end{equation}
With this, as a result of radiation accretion, the final mass of the PBHs at redshift $z_{i}\leq z_f \leq z_\mathrm{eq}$, can be written as:
\begin{equation}  \label{eq:Mfrad}
      M_{f,r}(z_f)  = \left[  \mathcal{C}\,\rho_{r,0} \, \left(\mathcal{I}(\Bar{z}_f) - \mathcal{I}(\Bar{z}_i)\right) + \frac{1}{M_i}\right]^{-1}\,.
\end{equation}

\begin{figure}[tb!]
\centering
\includegraphics[width = 0.6\textwidth]{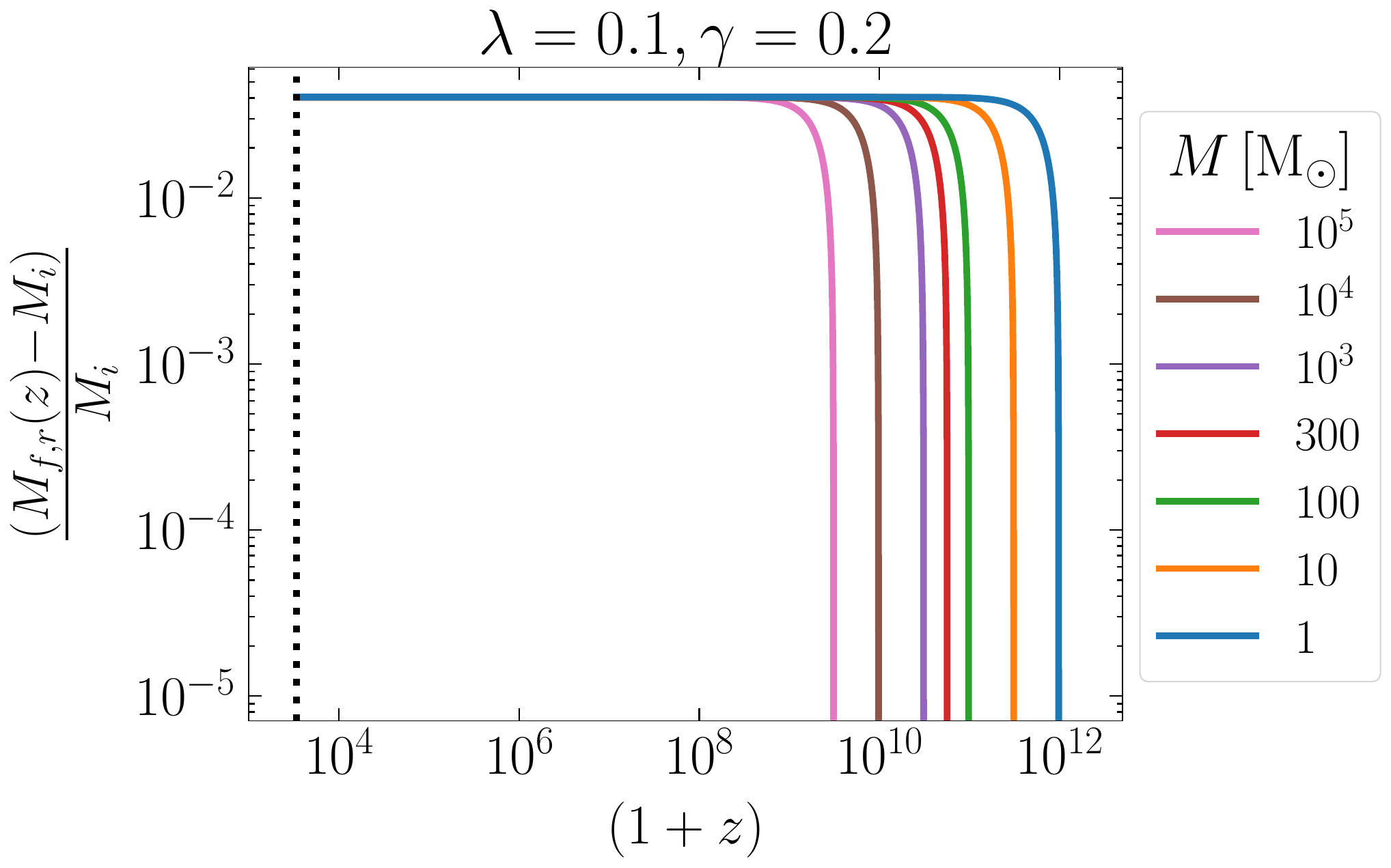}
\caption{Variation of the fractional change in the mass of the PBHs, as a function of $(1+z)$. Here, $M_i$ is the initial mass of the PBHs accreting radiation from their surroundings, resulting into their final mass as $M_{f,r}(z)$. The vertical black dotted line corresponds to the redshift of matter-radiation equality.}
\label{fig:racc2}
\end{figure}

The fractional change in the mass of the PBHs due to radiation accretion in RD is depicted in Figure~\ref{fig:racc2}.
This figure shows that for $\gamma = 0.2$ and accretion efficiency $\lambda = 0.1$, the mass of the PBHs due to radiation accretion can increase by a maximum of $4\%$. Moreover, the fractional change in the mass of the PBHs via radiation accretion appears to be independent of their initial mass. This is due to the fact that the formation of light PBHs takes place earlier and they start accreting when the density of surrounding radiation is high. Instead, more massive PBHs form later but have a larger Bondi radius $r_B \propto M$ and so can accrete as rapidly, even though the radiation background is more dilute. This ultimately results in the same growth for PBHs with different initial masses. We have also verified that the accreted mass grows with increase in the accretion efficiency $\lambda$, up to a maximum mass increase of $60\%$ for $\lambda = 1$.

Note that these results differ from what is claimed in Refs.~\cite{Nayak:2009wk,Nayak:2011sk}. They showed that for $\lambda = 0.1$, the mass of the PBHs due to accretion of radiation can increase up to $40\%$, around one order of magnitude larger than our estimates. This discrepancy arises because Refs.~\cite{Nayak:2009wk,Nayak:2011sk} ignore the factor $\gamma$ (the fraction of the Horizon mass which collapses to form the PBH), which leads to a delayed formation time for PBHs of fixed mass. In addition, in 
our work we use a more accurate expression for $z(t)$ at very early times, given in Eq.~\eqref{eq:zi}. Given that the accretion of radiation is dominated by early times when the radiation density is large, the final PBH mass is very sensitive to the precise time of PBH formation, leading to our smaller estimates compared to previous works.

While the accretion of radiation becomes inefficient by the time of matter-radiation equality, PBHs may still grow by accreting baryons at later times. In the following sections, we will therefore explore the mechanisms of baryonic accretion around PBHs.


\section{Bondi-Hoyle-Lyttleton (BHL) Accretion model for baryons}
\label{sec:Bondi-Hoyle-Lyttleton (BHL) Accretion model for baryons}
Now, we aim to provide a comprehensive study of the accretion of baryons around PBHs, evolving from radiation domination to the late Universe. We begin by revisiting the theoretical framework of the BHL model detailed in Refs.~\cite{Ricotti:2007au,DeLuca:2020bjf,DeLuca:2020fpg}, employing the speed of sound and PBH velocity, based on SPIK20~\cite{Serpico:2020ehh}. If PBHs do not make up all of the dark matter, then they can grow halos of particle DM around them. The additional gravitational potential due to these DM halos can substantially enhance the accretion rate, and so we extend the formalism to take this into account.\footnote{The influence of DM halos was not included in the previous section because PBHs have not had time to grow a substantial halo before radiation accretion becomes inefficient.} We use this formalism to determine the change in masses of the PBHs at low redshift $z\lesssim 15$. However, unlike ROM07~\cite{Ricotti:2007au}, we consider two possible scenarios for the late-time accretion around PBHs. In the first scenario, we consider that at all redshifts, PBHs accrete from the cosmological background fluid, designated as the linear regime (denoted by the subscript ``$\mathrm{L}$''). In the second scenario, we also take into account the potential infall of PBHs into virialized halos during structure formation in the late Universe ($z\leq 10$), designated as the non-linear regime (denoted by the subscript ``$\mathrm{NL}$'').

Assuming that the baryonic material in the Universe consists predominantly of hydrogen gas with mean number density~\cite{Ricotti:2007jk,Ricotti:2007au}:
\begin{equation}  \label{eq:ngas}
n_\mathrm{gas} =  200 \, \mathrm{cm}^{-3} \,  \left(\frac{ 1+ z }{1000}\right)^{3}\,,
\end{equation}
then the density of the gas can be written as:
\begin{equation}  \label{eq:density}
\rho_{b} = m_\mathrm{H} \cdot n_\mathrm{gas}\,,
\end{equation}
where $m_\mathrm{H} = 1.67 \times 10^{-27} \,\mathrm{kg}$. Then, using Eq.~\eqref{eq:Mdot}, the accretion rate of baryonic gas by a PBH of mass $M$ can be written as:
\begin{equation} \label{eq:MdotBHL} 
    \dot M_{\mathrm{BHL},0} = 4  \pi \lambda \,\rho_{b} \,v_\mathrm{eff} \,r_{\mathrm{B},0}^{2}\,,
\end{equation}
with $r_{\mathrm{B},0} = G M/v_\mathrm{eff}^{2}$ as the Bondi radius of isolated PBHs and $v_\mathrm{eff}$ as the effective velocity which depends on the speed of sound and the proper velocity of the PBHs. 
Here, the subscript ``$0$'' represents the accretion scenarios for \textit{isolated PBHs} (i.e.~PBHs without DM halos).

Following Refs.~\cite{Ricotti:2007jk,Ricotti:2007au}, we define the efficiency $\lambda$ for accretion from the cosmological background fluid as:
\begin{equation}  \label{eq:lambda}
 \lambda = \mathrm{exp} \left(\frac{9/2}{3 + \hat{\beta}^{0.75}}\right)\, x_\mathrm{cr}^{2} \,,
\end{equation}
with sonic radius 
\begin{equation}  \label{eq:sonicradius}
 x_\mathrm{cr} \equiv \frac{r_\mathrm{cr}}{r_{\mathrm{B},0}} = \frac{- 1 + \left(1 + \hat{\beta}\right)^{1/2}}{\hat{\beta}} \,,
\end{equation}
and gas viscosity
\begin{equation}  \label{eq:betahat}
\hat{\beta}  \equiv \left(\frac{M}{10^{4}\, \mathrm{M_{\odot}}}\right)\, \left(\frac{1 +z}{1000}\right)^{3/2} \, \left(\frac{v_\mathrm{eff}}{5.74\,\mathrm{km\,s^{-1}}}\right)^{-3} \left[0.257 + 1.45 \left(\frac{x_{e}}{0.01}\right) \left(\frac{ 1+ z }{1000}\right)^{5/2}\right] \,.
\end{equation}
Here, the gas viscosity $\hat{\beta}$ accounts for effects such as the cosmic expansion with recession velocity $v_\mathrm{H} = r_\mathrm{B}\cdot H(z)$~\cite{Ricotti:2007au} and Compton drag due to the scattering of free electrons (with electron fraction $x_e$) with CMB photons. The value of the accretion efficiency $\lambda$ given by Eq.~\eqref{eq:lambda} describes accretion from a uniform cosmic fluid of baryons. It should therefore be most accurate at high redshift, when the Universe is most homogeneous. However, at low redshift ($z \lesssim 10$), this description will eventually break down, as the local properties of the accretion flow becomes more relevant, during the process of structure formation.

Similar to Refs.~\cite{Tseliakhovich:2010bj,Poulin:2017bwe,Ali-Haimoud:2016mbv,Serpico:2020ehh}, we consider the speed of sound as:
\begin{equation}  \label{eq:c_serpico} 
c_\mathrm{s} \simeq 6\,  \mathrm{km\,s^{-1}}    \: \left(\frac{1 + z}{1000}\right)^{1/2}   \quad ; \quad 100 \lesssim z \lesssim 1000\,,
\end{equation}
and the proper velocity of the PBHs as:
\begin{equation}  \label{eq:vpbhlinear} 
v_\mathrm{pbh,L} \simeq  \mathrm{min} \left[1, \frac{1 + z}{1000}\right] \cdot \: 30\:\, \mathrm{km\,s^{-1}} \,.
\end{equation}
The formation of structure at low redshift may lead PBHs to become bound in virialized halos, with much larger typical velocities. So, we redefine the proper velocity of the accreting PBHs as~\cite{Serpico:2020ehh,Ricotti:2007au}:
\begin{equation} \label{eq:vpbhfull} 
v_{\mathrm{pbh}, \mathrm{total} }   = 
\begin{dcases}
       \text{$v_\mathrm{pbh,L}$,}   &   \quad   \text{$z > 10$}\\
       \text{$v_\mathrm{pbh,NL}$,}  &   \quad   \text{$z \leq 10$}\\
\end{dcases}\,\;,
\end{equation}
where
\begin{equation}  \label{eq:vpbhnonlinear} 
v_\mathrm{pbh,NL} \simeq  17\, \mathrm{km\,s^{-1}} \cdot \left(\frac{M(z)}{10^{8}\, \mathrm{M_{\odot}}}\right)^{1/3} \left(\frac{1+z}{10}\right)^{1/2} \,,
\end{equation}
is the velocity of PBHs in $2\sigma$ density perturbations, estimated using the Press-Schechter formalism~\cite{William1974}, with
\begin{equation}  \label{eq:masssigma} 
M(z) = 8.8\times 10^{12} \cdot \mathrm{exp}\left[-1.8\,(1+z)\right]\,,
\end{equation}
being the mass of such density perturbations~\cite{Ricotti:2007au}. The infall of PBHs into virialized halos during structure formation is a probabilistic process so not all PBHs will become bound in structure at low-redshift. However, as we detail in Appendix~\ref{sec:Fraction of PBHs in virialized halos}, by $z = 0$ around $75\%$ of PBHs will be inside structures whose virial velocity is larger than the corresponding value of $v_{\mathrm{pbh,L}}$, emphasizing the relevance of taking into account these non-linear velocities.
\begin{figure}[tb!]
\centering
\includegraphics[width = 0.45\textwidth]{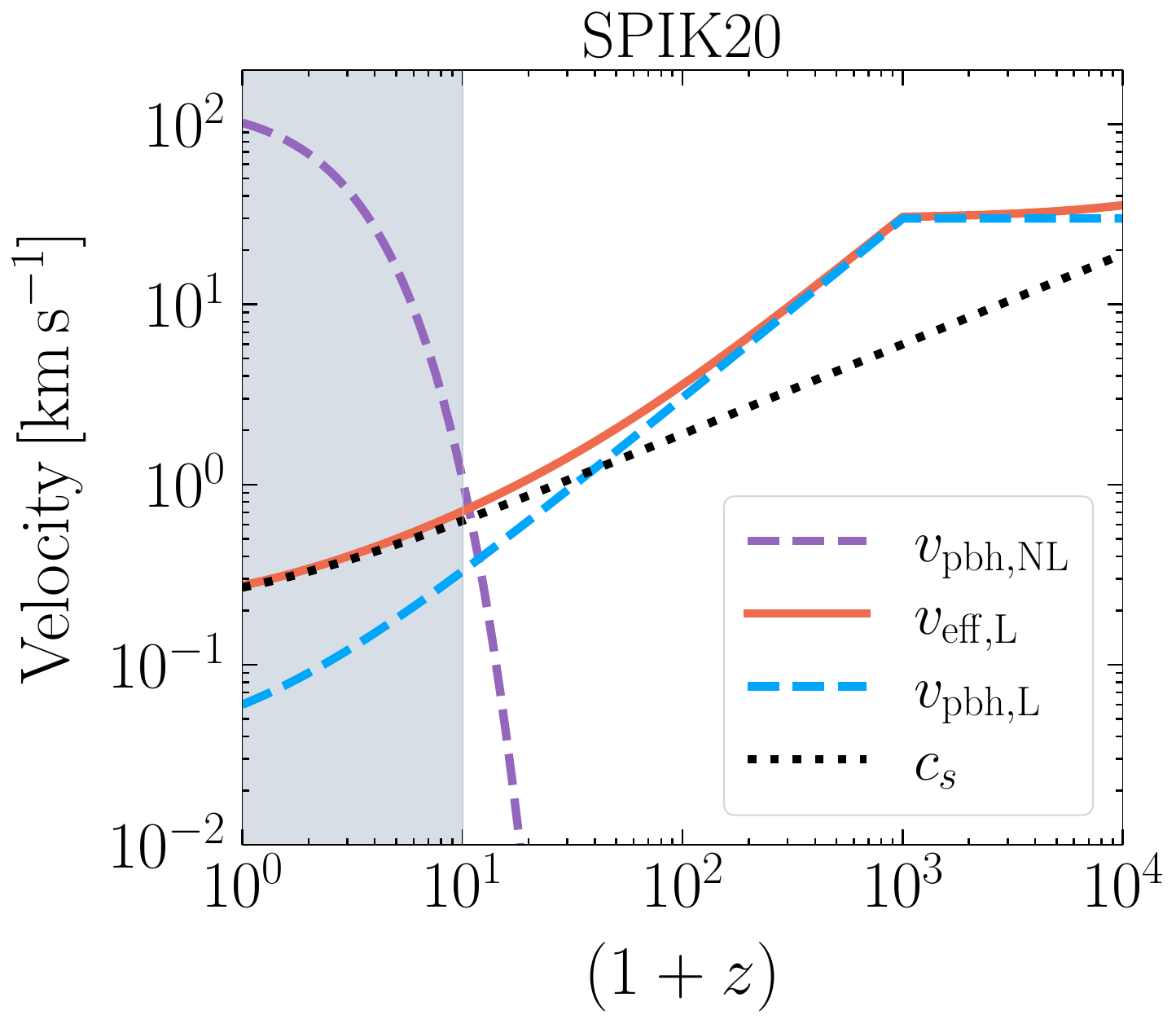}
\caption{Variation of the speed of sound $c_s$, proper velocity $v_\mathrm{pbh}$ of the PBHs and effective velocity $v_\mathrm{eff}$, as a function of redshift $(1+z)$. Here, the values of $v_\mathrm{eff,L}$ and $v_\mathrm{eff,NL}$ are evaluated using $v_\mathrm{eff} = \sqrt{v_\mathrm{pbh}^{2} + c_s^{2}}$~\cite{Poulin:2017bwe,Serpico:2020ehh,Ali-Haimoud:2016mbv,Tseliakhovich:2010bj}. The grey shaded region shows the non-linear regime corresponding to the era where PBHs fall into virialized halos during structure formation.}
\label{fig:velSPIK20}
\end{figure}

Figure~\ref{fig:velSPIK20} shows the variation of the speed of sound and the proper velocity of the PBHs given by Eqs.~\eqref{eq:c_serpico} and~\eqref{eq:vpbhfull} (which we refer to as the SPIK20 velocity profiles). At high redshift, the velocity of the PBHs dominates over the speed of sound, and as the PBH velocities are redshifted away, they eventually become subdominant. 
As mentioned earlier, in the non-linear regime corresponding to $z \leq 10$, the $v_\mathrm{pbh}$ increases drastically, leading to a significant increase in the effective velocity too.

\subsection{Accretion around PBHs with DM Halos}
\label{sec:Accretion around PBHs with DM halos}
Among other constraints, the observations of the LVK Collaboration constrain the abundance of PBHs in cold dark matter (CDM) to be less than $f_\mathrm{pbh} \sim 10^{-3} - 10^{-2}$ in the mass range $1-100\,M_\odot$~\cite{Sasaki:2016jop,Raidal:2018bbj,DeLuca:2020qqa,Carr:2020gox}. So, if PBHs do not make up all of the dark matter then other particle candidates of dark matter may exist in the Universe. In that case, such dark matter particles can become gravitationally bound to the PBHs leading to the formation of dense DM mini-halos around them. To avoid the self-depletion of DM halos (along with associated gamma-ray signatures), we consider scenarios where DM particles are not WIMP-like, i.e.\ they have a very small annihilation cross-section~\cite{Mack:2006gz,Ricotti:2007jk,Lacki:2010zf,Cerdeno:2010jj,Boucenna:2017ghj,Bertone:2019vsk,Carr:2020mqm}. The presence of these DM particles around PBHs enhances the gravitational potential leading to enhanced accretion of surrounding material. To quantify this effect, we re-examine the framework of baryonic accretion around PBHs with DM halos, mentioned in Refs.~\cite{Ricotti:2007au,DeLuca:2020bjf,DeLuca:2020fpg}. In this study, we also incorporate the generalized modification of Bondi radius and the corrections in PBHs velocity in the late Universe.

After kinetic decoupling, DM particles start to stream freely in the Universe which can lead to the formation of DM halos around PBHs. For spherical symmetry, the density of the DM halos existing around isolated PBHs is given by~\cite{1984ApJ...281....1F,1985ApJS...58...39B,Adamek:2019gns,Boudaud:2021irr,Jangra:2023mqp}:
\begin{equation} \label{eq:DMdensity}
  \rho_{h} (r) = \Bar{\rho}\, \left(\frac{r}{r_\mathrm{ta}}\right)^{-\alpha}\,,
\end{equation}
with $\alpha = 9/4$ and
\begin{equation} \label{eq:rhobar}
 \Bar{\rho} = \frac{\Omega_\mathrm{cdm}}{\Omega_\mathrm{m}} \,\frac{\rho_\mathrm{eq}}{2}\, (2 G M t_\mathrm{eq}^{2})^{3/4} \,r_\mathrm{ta}^{-9/4}\,.
\end{equation}
Here, $r_\mathrm{ta}$ represents the size of the DM halo, also known as the turnaround radius defined at the turnaround time $t_\mathrm{ta}$~\cite{Adamek:2019gns,Jangra:2023mqp}:
\begin{equation} \label{eq:turnaroundradius}
  r_\mathrm{ta} = (2 G M t_\mathrm{ta}^{2})^{1/3}\,,
\end{equation}
illustrating the continuous growth of the halo. 
Moreover, $\rho_{\mathrm{eq}}$ denotes the total energy density at matter-radiation equality ($z_\mathrm{eq} \approx 3400$) and $\Omega_\mathrm{cdm}/\Omega_\mathrm{m} \approx 0.85$ indicates the fraction of CDM in the total matter density of the Universe~\cite{Planck:2015fie}. The typical size of the DM halos lies in range of $\mathrm{kpc}-\mathrm{Mpc}$ and the Bondi radius typically ranges from $\mathrm{km}-\mathrm{AU}$. So, the size of DM halos can be many orders of magnitude larger than the Bondi radius. Also, in general, the simultaneous accretion of DM particles and baryons might overlap with each-other, leading to very complex dynamics. But for simplicity, we assume that the accretion of baryons around PBHs and the growth of DM halos are two independent processes.

From Eq.~\eqref{eq:DMdensity}, the mass of the DM halos for spherical symmetry can be calculated as:
\begin{equation} \label{eq:halomass}
 M_{h}(r) = 
    \frac{4\pi \Bar{\rho} }{ (3 - \alpha)}\,\begin{dcases}
       r^{3-\alpha} \,r_\mathrm{ta}^{\alpha} & \quad \text{for } r<r_\mathrm{ta}\,,\\
       r_\mathrm{ta}^{3} &\quad\text{for } r \ge r_\mathrm{ta} \,.
     \end{dcases}
 \end{equation}
Then, using the reference point at infinity, the gravitational potential due to the DM halo can now be computed as:
\begin{equation} \label{eq:phi}
     \Phi_{h}(r) = -G \int_{r}^{\infty} \frac{M_{h}(r) }{r^{2}} \,\mathrm{d}r\,,
 \end{equation}
leading to:
\begin{equation} \label{eq:phieff} 
     \Phi_{h}(r) = \begin{dcases}
       \frac{-GM_{h}}{(p-1 )}\left[\frac{p}{r_\mathrm{ta}}- \frac{r^{p-1}}{r_\mathrm{ta}^{p}}\right] &\quad\text{for } r<r_\mathrm{ta}\,,\\
       -\frac{GM_{h}}{r_\mathrm{ta}} &\quad\text{for } r \ge r_\mathrm{ta} \,.
     \end{dcases}
 \end{equation}
where $ M_{h} \equiv M_{h}(r_\mathrm{ta}) = \left(4\pi \Bar{\rho}/ (3 - \alpha)\right)r_\mathrm{ta}^{3}$ and $p = (3 - \alpha)$.

The Bondi radius separates regions of sub-sonic and super-sonic flow. Assuming the infall of baryonic material from infinity, this implies that $c_s^2 = v_\mathrm{esc}^2(r_B)$. Taking into account the relative motion of the PBH ($c_s^2 \rightarrow v_\mathrm{eff}^2$), and in the absence of DM halos around the PBHs, this expression leads to:
\begin{equation} \label{eq:6}
    v_\mathrm{eff}^{2} =  \frac{G M}{r_{\mathrm{B},0}}\,.
\end{equation}
In the presence of a DM halo around the PBH with a density profile of $\rho(r) \propto r^{-\alpha}$ and total mass $M_{h}$, we can generalize the above equation as~\cite{Serpico:2020ehh}:
\begin{equation} \label{eq:7}
    v_\mathrm{eff}^{2} =  \frac{G M}{r_\mathrm{B,eff}} \: - \: \Phi_{h}(M_h, r_\mathrm{B,eff}, z )\,,
\end{equation}
where we call $r_\mathrm{B,eff}$ the \textit{effective} Bondi radius.  
Now, substituting $\Phi_{h}$ given by Eq.~\eqref{eq:phieff}, we can rewrite the above equation as:
\begin{equation} \label{eq:8}
    v_\mathrm{eff}^{2} =  \frac{G M}{r_\mathrm{B,eff}} \: + \: \frac{G M_{h}}{r_\mathrm{B,eff}} \left\{\Theta\left(r_\mathrm{B,eff} - r_\mathrm{ta}\right) + \frac{\Theta\left(r_\mathrm{ta} - r_\mathrm{B,eff}\right)}{1-p}\left[\left(\frac{r_\mathrm{B,eff}}{r_\mathrm{ta}}\right)^{p} - p \,\left(\frac{r_\mathrm{B,eff}}{r_\mathrm{ta}}\right)  \right]\right\}\,,
\end{equation}
leading to:
\begin{equation} \label{eq:9}
    r_\mathrm{B,eff} =  \frac{G \left(M + M_{h}\right)}{v_\mathrm{eff}^{2}}\,;\quad \text{if }  r_\mathrm{ta}<r_{\mathrm{B},h} \,,
\end{equation} 
and 
\begin{equation} \label{eq:rBeffBHL} 
   \frac{r_{\mathrm{B},h}}{(1 - p)} \cdot \frac{1}{r_\mathrm{ta}^{p}} \cdot r_\mathrm{B,eff}^{p} \,-\, \left(1 + \frac{p}{(1 - p)} \cdot \frac{r_{\mathrm{B},h}}{r_\mathrm{ta}}\right) \cdot  r_\mathrm{B,eff} \,+ \,r_{\mathrm{B},0} = 0 \,; \quad \text{if }  r_\mathrm{ta} > r_{\mathrm{B},h} \,,
\end{equation}
with $r_{\mathrm{B}, h} = G M_{h}/v_\mathrm{eff}^{2}$ being the Bondi radius of the DM halo considered as a point mass. Equation~\eqref{eq:rBeffBHL} can then be solved numerically for  $r_\mathrm{B,eff}$.

\subsection{Accretion rates}
\begin{figure}[tb!]
\centering
\includegraphics[width = 0.38\textwidth]{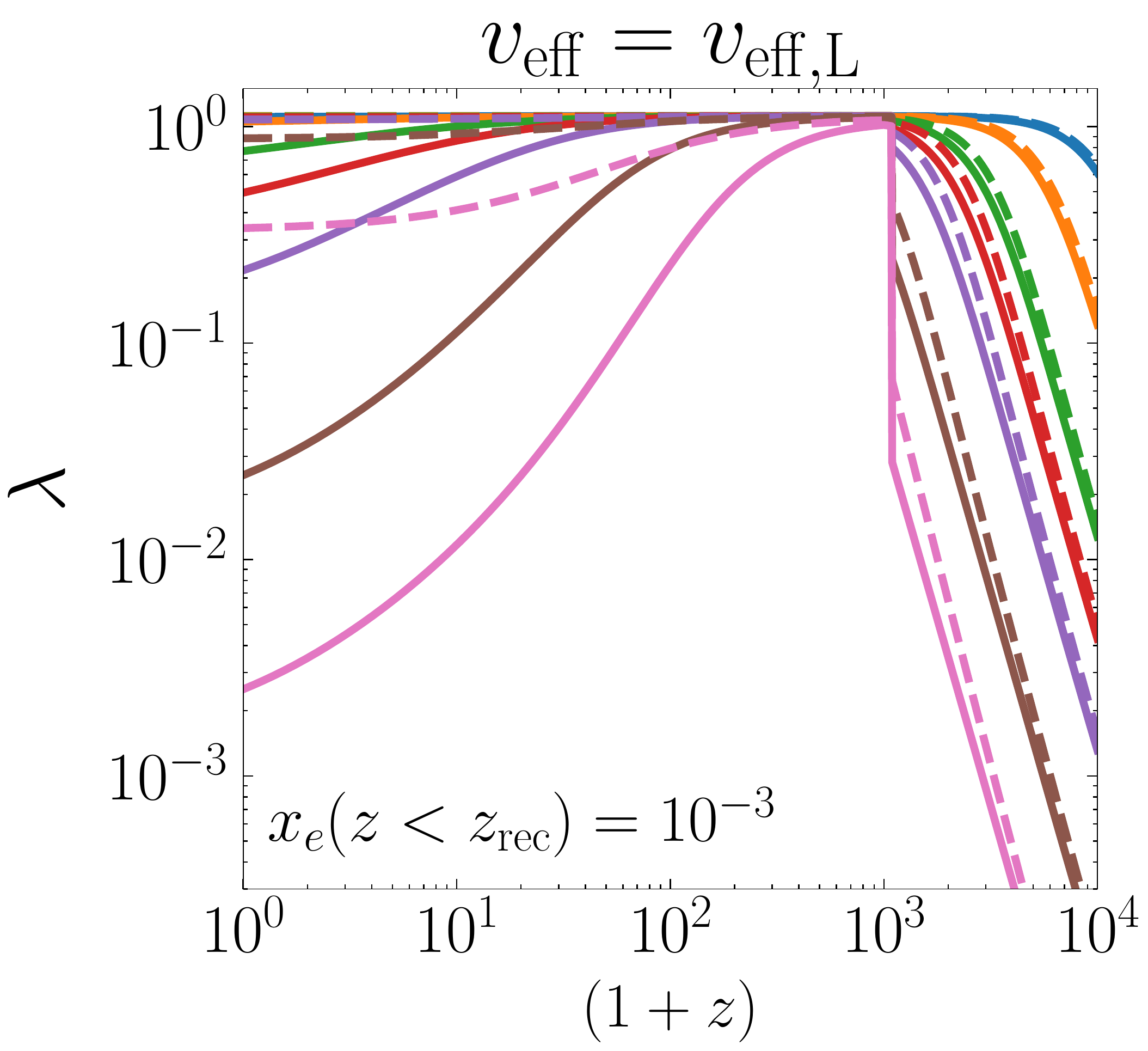}
\hspace*{\fill}
\includegraphics[width = 0.6\textwidth]{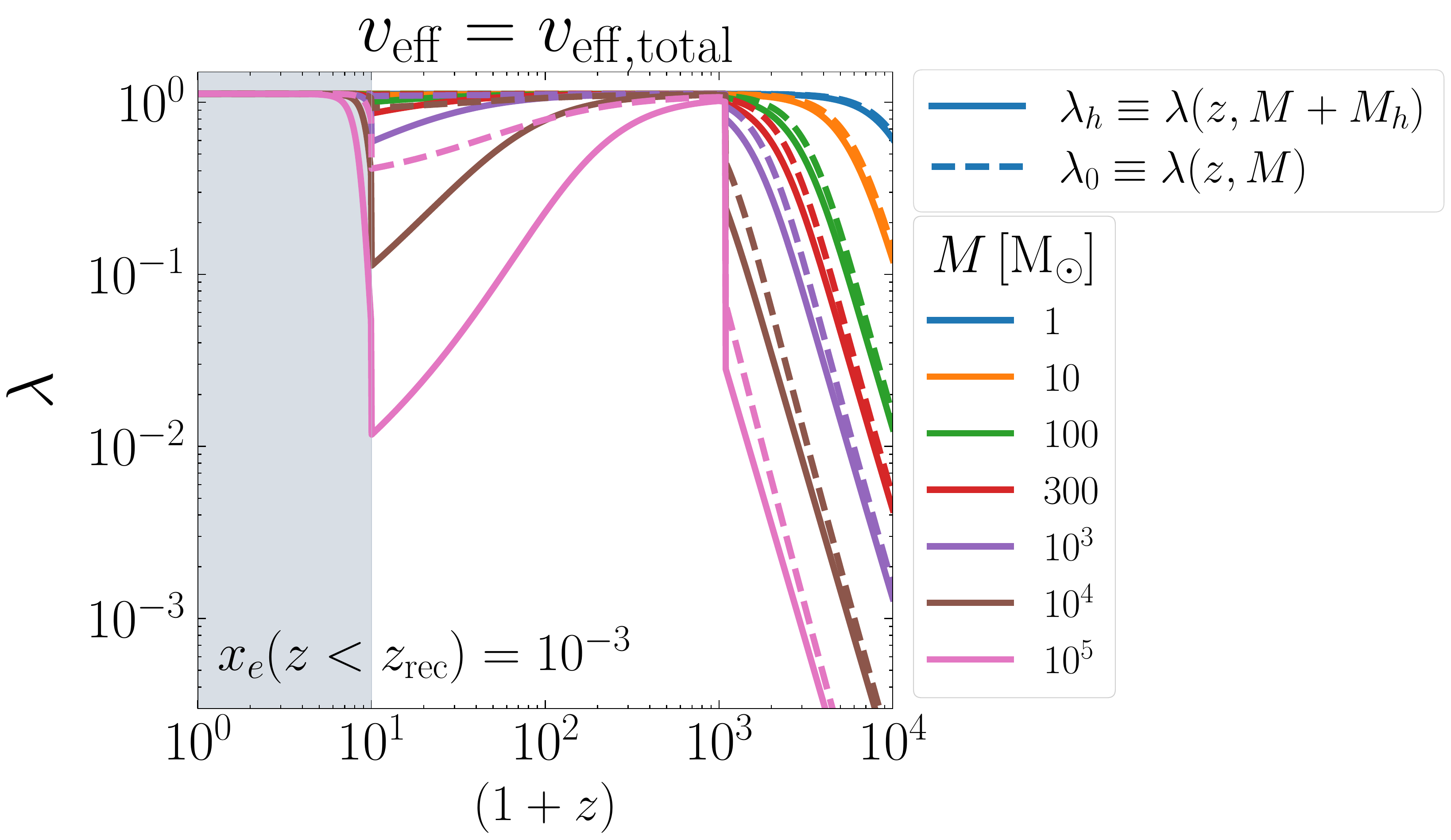}
\caption{Variation of the accretion efficiency $\lambda$ with redshift $(1+z)$. The left panel illustrates the scenarios of PBHs evolving under the Hubble expansion only while the right panel also accounts for the effects of late structure formation on the motion of PBHs (as shown by the grey shaded region). Here, the DM halos around the PBHs possess the density profile of $\rho(r) \propto r^{-9/4}$ and for $z \leq z_\mathrm{rec} \approx 1089$, the value of constant electron fraction is considered as $x_e = 10^{-3}$~\cite{Ricotti:2007au}.}
\label{fig:lambda}
\end{figure}

In Figure~\ref{fig:lambda}, we show the variation of the accretion efficiency $\lambda$ with redshift, based on the velocity profiles shown in Figure~\ref{fig:velSPIK20}. The dashed colored lines represent the values of $\lambda$ for isolated PBHs while the solid colored lines correspond to the counterpart scenarios of PBHs with DM halos.\footnote{In the presence of DM halos, we calculate $\lambda$ assuming an accreting point mass of $(M + M_h)$. While calculating the accretion rate, the extended nature of the halo is taken into account through the effective Bondi radius, as described in Sec.~\ref{sec:Accretion around PBHs with DM halos}.} At high redshift, the Universe is fully ionised ($x_e = 1$) meaning that Compton drag from the scattering of free electrons off CMB photons leads to a large gas viscosity and therefore a small accretion efficiency. As the Universe becomes more dilute, the effects of Compton drag and the contribution from the Hubble expansion decrease, leading to an increase in the accretion efficiency. At recombination, the number of free electrons drops rapidly (we assume $x_e = 10^{-3}$) due to which the Compton drag declines abruptly, typically causing the accretion efficiency to saturate at $\lambda \approx 1$.
Then, for $z < z_\mathrm{rec}$, the decreasing PBH velocity leads to an increase in the Bondi radius, leading to a suppression of the accretion efficiency as the recession velocity of the gas $v_\mathrm{H} \propto r_\mathrm{B}$ is enhanced. Similarly, we note that the accretion efficiency $\lambda$ also decreases as we increase the PBH mass, due to an increase Bondi radius and therefore gas viscosity. Including DM halos (solid lines) leads to a further increase in the Bondi radius, suppressing the accretion efficiency with respect to isolated PBHs (dashed lines), with the difference becoming larger at low redshift with the growing mass of the DM halos.

In the presence of DM halos, the baryonic accretion rate of PBHs gets modified as:
\begin{equation} \label{eq:MdotBHLh}
  \dot M_{\mathrm{BHL},h}  = 4 \pi \lambda\,  \rho_{b} \, v_\mathrm{eff}\, r_\mathrm{B,eff}^{2}\,.
\end{equation}
Here, the subscript ``$h$'' indicates the presence of DM halos around isolated PBHs. During the accretion process, as the accreted material gets closer to the PBHs, its energy is converted into heat which can be partially converted into radiation. The Eddington accretion rate is considered as a limiting case, in which the resulting radiation pressure balances the gravitational force on the infalling material. For an isolated PBH of mass $M$, the Eddington rate is given by~\cite{1926ics..book.....E,Ricotti:2007au}:
\begin{equation}  \label{eq:MdotEDD}
 \dot M_\mathrm{Edd} = 1.44\times 10^{17} \, \left(\frac{M}{M_{\odot}}\right)\, \mathrm{g\,s^{-1}}\,,
\end{equation}
with which we define the dimensionless accretion rate:
\begin{equation}  \label{eq:mdot}
 \dot m =  \frac{\dot M}{\dot M_\mathrm{Edd}} \,.
\end{equation}

\begin{figure}[tb!]
\centering
\includegraphics[width = 0.4\textwidth]{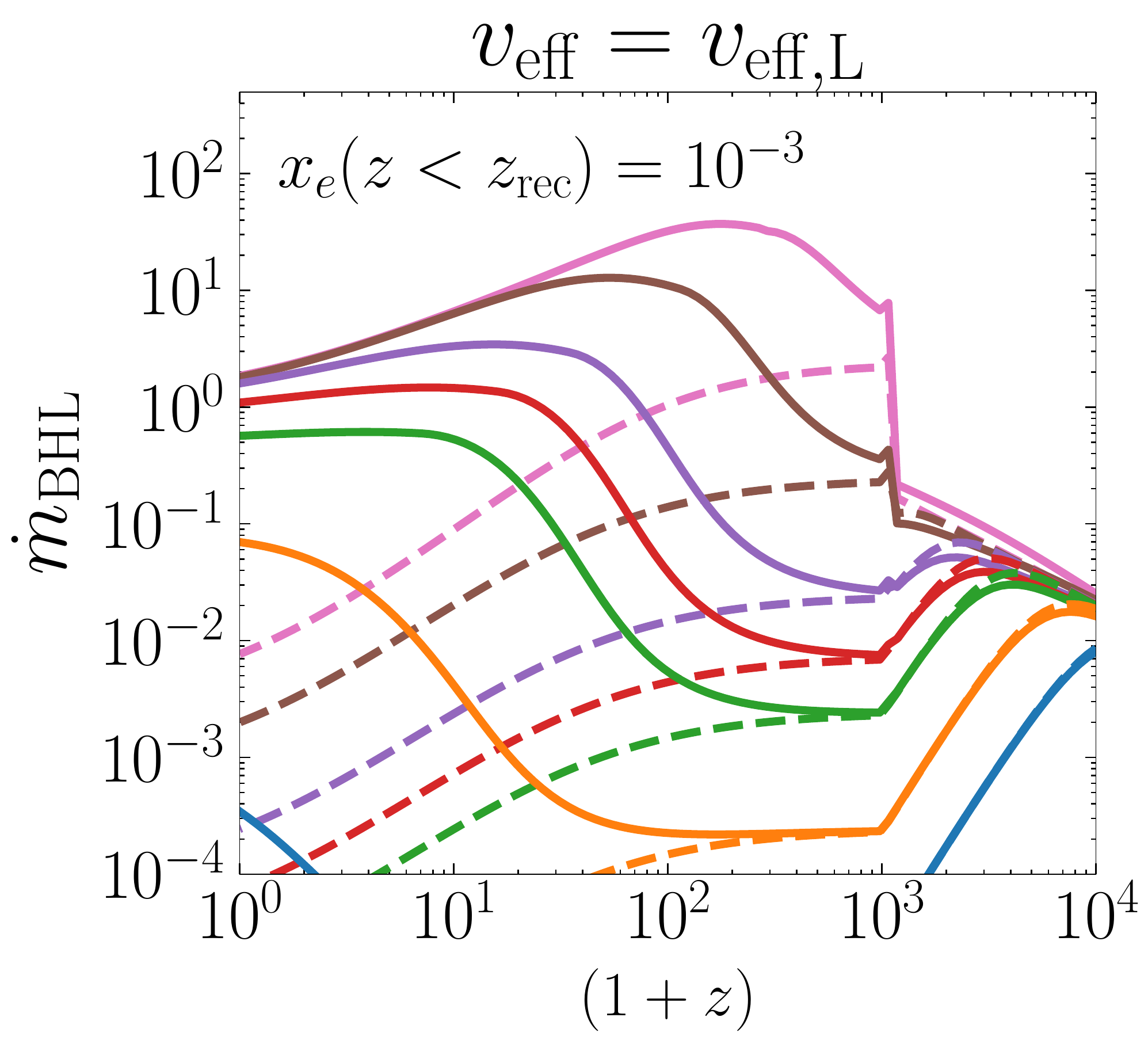}
\hspace*{\fill}
\includegraphics[width = 0.568\textwidth]{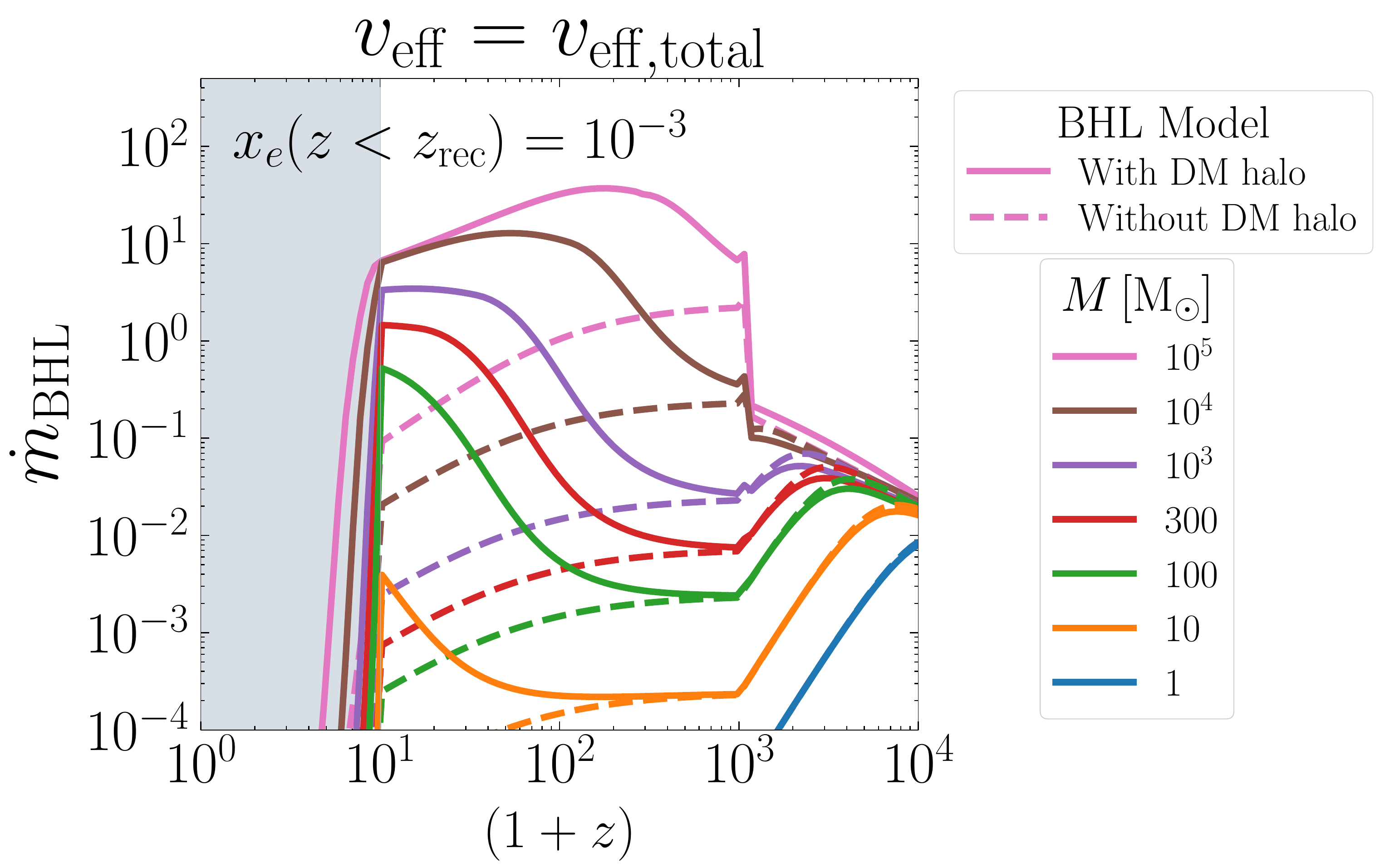}
\caption{Variation of the dimensionless accretion rate for PBHs with and without DM halos, as a function of the redshift $(1+z)$. Both the panels correspond to the baryonic accretion around PBHs with constant electron fraction $x_e = 10^{-3}$, post recombination. The grey shaded region in the right panel depicts the accretion rates for PBHs falling into virialized halo during structure formation.}
\label{fig:mdotBHL}
\end{figure}

\begin{figure}[tb!]
\centering
\includegraphics[width = 0.49\textwidth]{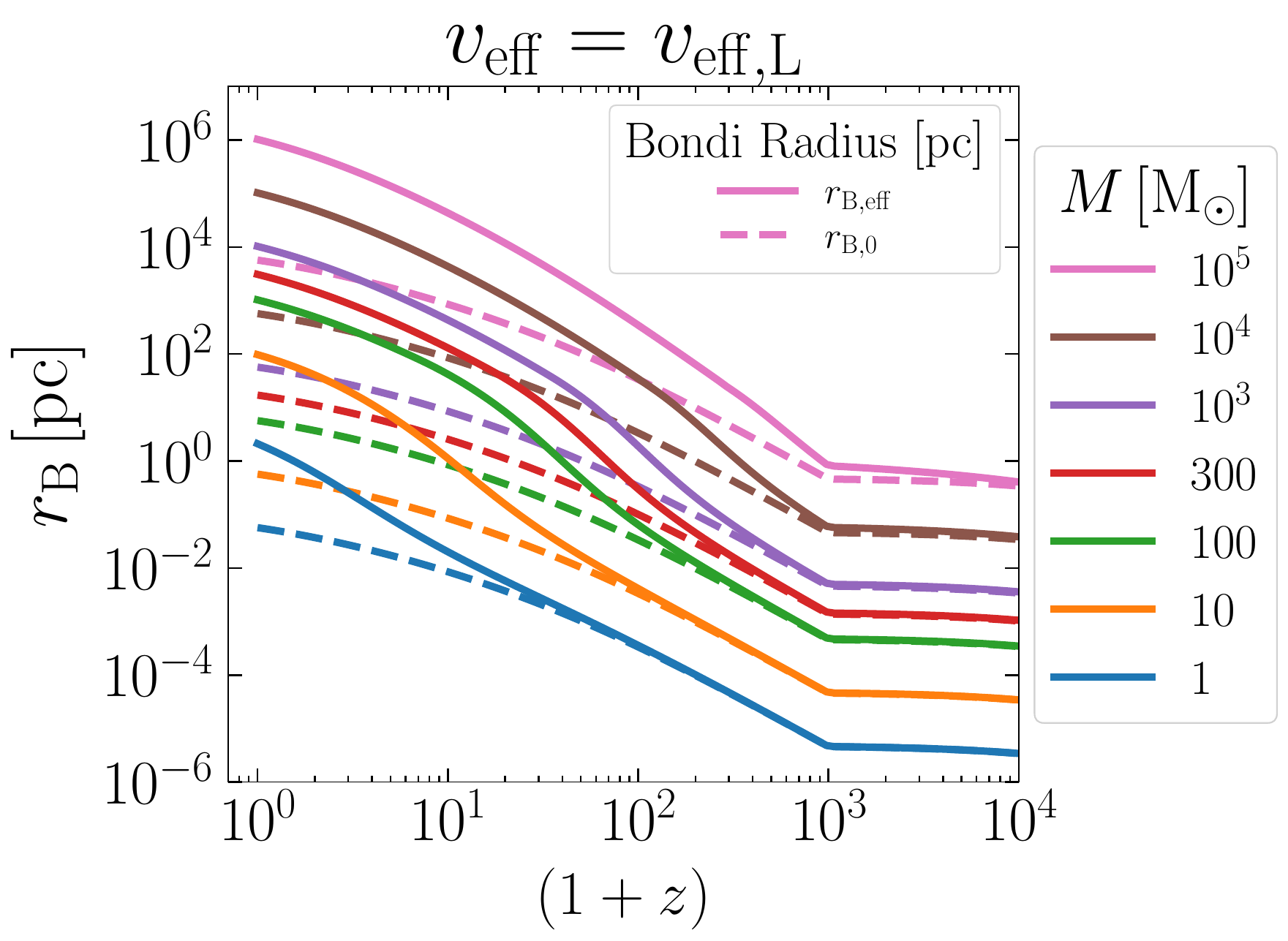}
\hfill
\includegraphics[width = 0.49\textwidth]{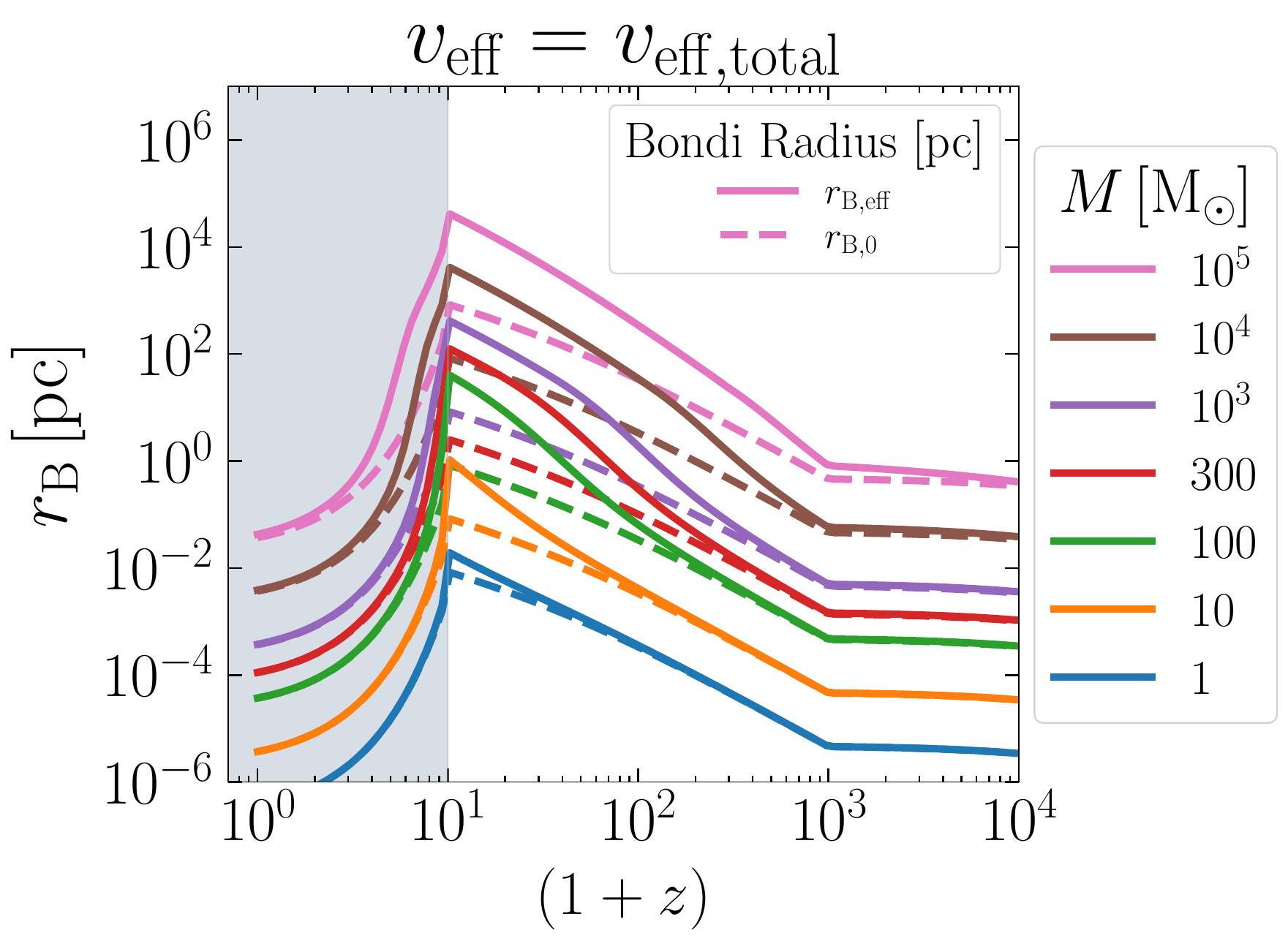}
\caption{Variation of the Bondi radius in BHL model for PBHs with and without DM halos. The DM halos around the PBHs share the density profile of $\rho_{h}(r) \propto r^{-9/4}$~\cite{Adamek:2019gns,Jangra:2023mqp}. The left panel shows the Bondi radius of PBHs evolving under Hubble expansion while the left panel depicts the corresponding values by including the effects of the fall of PBHs into virialized halos for $z \leq 10$.}
\label{fig:BondiradiusSPIK20}
\end{figure}

The variation of these dimensionless accretion rates of PBHs with and without DM halos in the BHL model is illustrated in Figure~\ref{fig:mdotBHL}.
Both the panels of this figure demonstrate that the accretion rates increase with increase in masses of the PBHs, corresponding to a stronger gravitational potential for capturing gas. While the accretion rate in Eq.~\eqref{eq:MdotBHLh} scales naively as $r_{B}^{2} \sim M^{2}$, the accretion efficiency decreases with $M$, as we saw in Fig.~\ref{fig:lambda}. This means that overall the dimensionless accretion rate $\dot{m}$ scales roughly $\propto M$.
At high redshift, the accretion rates of PBHs with and without DM halos are comparable, as the DM halos have not yet had sufficient time to grow substantially. At lower redshift, the accretion rates of PBHs with DM halos are larger by $\approx \mathcal{O}(100)$ compared to scenarios without DM halos. This is attributed to the growth of DM halos at lower redshift resulting in $r_\mathrm{B,eff}(M + M_{h}) \gg r_{\mathrm{B},0} (M)$, as illustrated in Figure~\ref{fig:BondiradiusSPIK20}. However, this enhancement diminishes notably for PBHs falling into virialized halos during structure formation (for $z \leq 10$) which results from the suppression in the values of Bondi radius (right panel of Figure~\ref{fig:BondiradiusSPIK20}). The sudden rise in the accretion rate at $z = z_\mathrm{rec}$ which is seen in Fig.~\ref{fig:mdotBHL} stems from the rise in the values of accretion efficiency $\lambda$ due to the shift in electron fraction $x_e$ from $1$ or $10^{-3}$.

So, Figure~\ref{fig:mdotBHL} clearly demonstrates that the presence of DM halos can notably enhance the baryonic capture around PBHs. However, these enhancements completely diminish when PBHs fall into the virialized halos around $z \leq 10$, underscoring the intricate interplay between the velocity of PBHs and accretion dynamics in cosmological scenarios. In the next section, we extend our analysis using a comparatively refined model known as the Park-Ricotti (PR) accretion model. This model will account for the effects of radiative feedback to more accurately capture the dynamics of gas accretion around PBHs with and without DM halos.

\section{Park Ricotti (PR) Accretion model for baryons}
\label{sec:Park Ricotti (PR) Accretion model}
\begin{figure}[tb!]
\centering
\includegraphics[width = 0.55\textwidth]{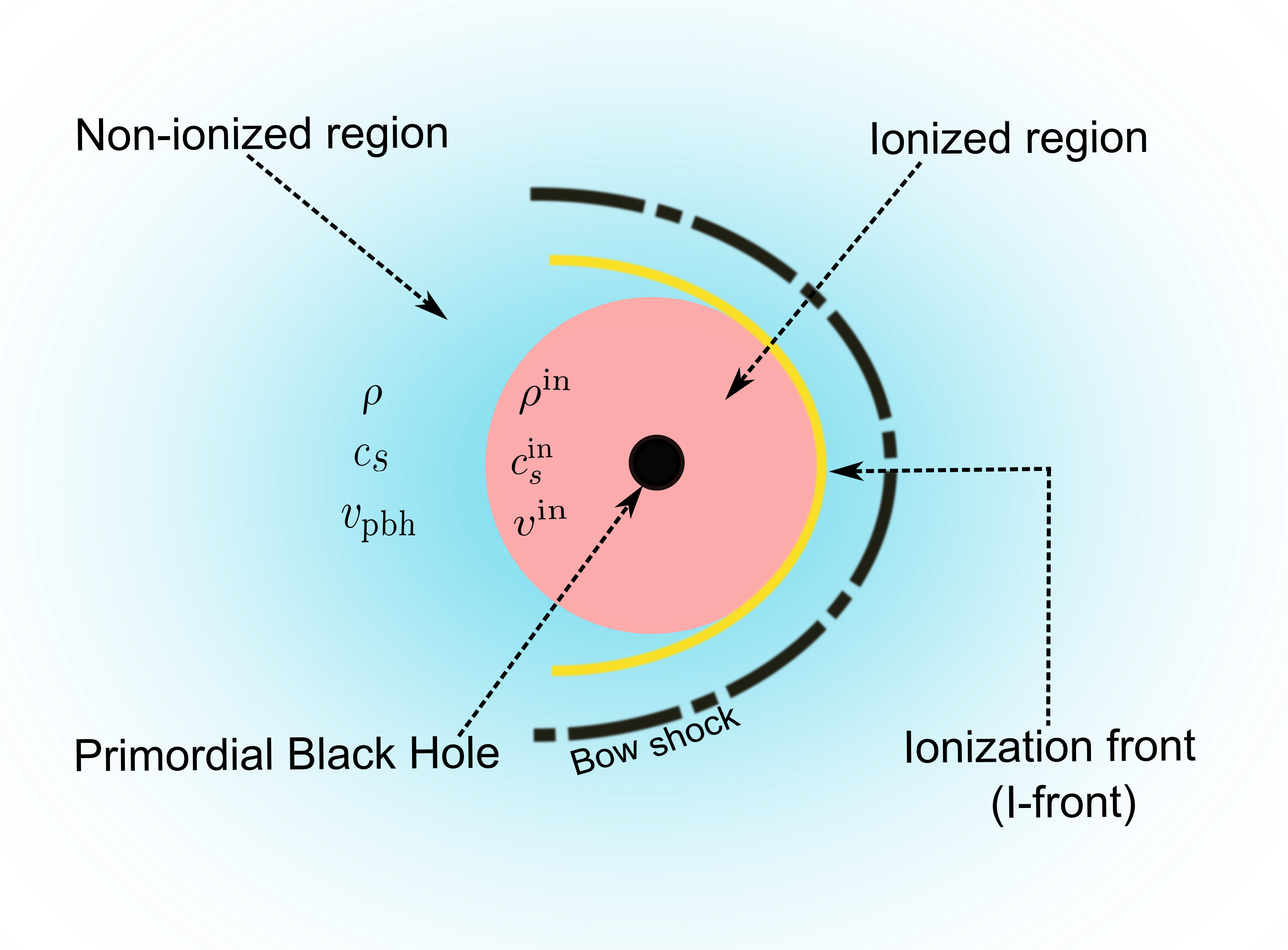}
\caption{Accretion in the Park-Ricotti (PR) model. The blue shaded region denotes the non-ionized gas surrounding the accreting PBH, while the light red shaded region represents the ionized gas produced by radiative feedback. The boundary between the ionized and the neutral gas is referred to as the ionization front (I-front), shown by the yellow colored circle, near to which a dense layer of further incoming neutral gas gets formed. Moreover, the supersonic motion of PBHs may compress the gas leading to the formation of a  bow shock (shown by the black colored dash-dotted circle) which creates outward thermal pressure and further reduces the accretion rate.}
\label{fig:PRmodel}
\end{figure}

The conventional BHL accretion model (discussed in Sec.~\ref{sec:Bondi-Hoyle-Lyttleton (BHL) Accretion model for baryons}) typically neglects radiative feedback on the accreting material. References~\cite{Park:2010yh,Park:2011rf,Park:2012cr} explored the impact of this important process in the spherical accretion of baryons around isolated and stationary intermediate mass black holes (IMBHs) and provided a formalism which we will refer to as the Park-Ricotti (PR) model. In this model, the emitted radiation (specifically UV and X-rays) ionizes the gas within the Bondi radius, forming a hot ionized region. As illustrated in Fig.~\ref{fig:PRmodel}, this hot bubble forms a boundary between the ionized and the neutral gas, known as the ionization front (I-front). The outward pressure of the gas within the ionized region hinders the further inward flow of baryons and significantly reduces the accretion rates. The PR formalism consists of an analytical prescription for accretion (summarised below), validated using numerical simulations, which showed that the average accretion rates of IMBHs are substantially smaller than the corresponding values predicted by the traditional BHL model. This model has previously been applied to the detectability of Galactic astrophysical black holes~\cite{Scarcella:2020ssk} and to the imprints of PBH accretion in the Cosmic Microwave Background~\cite{Agius:2024ecw}.

Following the methodology outlined in Refs.~\cite{Park:2012cr,Scarcella:2020ssk}, for an isolated PBH of mass $M$ (without DM halo), the accretion rate within the ionized region in the Park-Ricotti (PR) model can be expressed as:
\begin{equation} \label{eq:MdotPR} 
  \dot  M_{\mathrm{PR},0}  = 4 \pi \, \rho^\mathrm{in} \, v_\mathrm{eff}^\mathrm{in}\, {r_{\mathrm{B}, 0}^\mathrm{in}}^{2}\,,
\end{equation}
with
\begin{equation} \label{eq:25}
  {r_{\mathrm{B}, 0}^\mathrm{in}} = \frac{G M}{ {v_\mathrm{eff}^\mathrm{in}}^{2}}\,,
\end{equation}
and $v_\mathrm{eff}^\mathrm{in} = \sqrt{{v^\mathrm{in}}^{2} + {c_\mathrm{s}^\mathrm{in}}^{2}}$. Here, the subscript ``$\mathrm{in}$'' signifies the region ionized by the trapped radiation, in which the rate of accretion as per BHL model is considered to be valid~\cite{Scarcella:2020ssk}. Note that the size of the ionized region is expected to be always $\mathcal{O}(100)$ times larger than the Bondi radius, such that we can apply the BHL formalism there~\cite{Sugimura:2020rdw}. The value of  $c_\mathrm{s}^\mathrm{in}$ depends on the details of radiative feedback inside the bubble and is typically treated as a free parameter. Analogous to Ref.~\cite{Scarcella:2020ssk}, we fix $c_\mathrm{s}^\mathrm{in} = 25 \, c_\mathrm{s}$, though in Sec.~\ref{sec:PR_accretion_rates} we comment on how varying this constant affects the accretion rates. In contrast to the standard BHL model, which accounts for the interaction with the surroundings indirectly through the accretion efficiency $\lambda$, a similar factor can be included in the PR Model to incorporate interactions like coupling with the background radiation and Hubble expansion. However, we assume that the dynamics and properties of the gas in the ionized region is captured by the value of $c_\mathrm{s}^\mathrm{in}$ and so we set the accretion efficiency to unity.

Fixing $c_\mathrm{s}^\mathrm{in} = 25 \, c_\mathrm{s}$, PBH relative velocity inside the ionized region $v^\mathrm{in}$ can be determined from its counterpart in the surrounding medium through conservation of mass and momentum, as~\cite{Scarcella:2020ssk}:
\begin{equation} \label{eq:vin} 
v^\mathrm{in}    = 
\begin{dcases}
       \frac{\rho}{\rho^\mathrm{in}}\, v_\mathrm{pbh} &\quad\text{for } v_\mathrm{pbh}\geq v_\mathrm{R}\,,\\
       c_\mathrm{s}^\mathrm{in} & \quad \text{for } v_\mathrm{D} < v_\mathrm{pbh} < v_\mathrm{R}\,,\\ 
       \frac{\rho}{\rho^\mathrm{in}}\, v_\mathrm{pbh} & \quad \text{for } v_\mathrm{pbh}\leq v_\mathrm{D}\,.
\end{dcases}
\end{equation}
Here, $v_\mathrm{D}$ and $v_\mathrm{R}$ are the roots of:
\begin{equation}
\Delta = \left(v_\mathrm{pbh}^{2} + c_\mathrm{s}^{2}\right)^{2} - 4 \,v_\mathrm{pbh}^{2}\,c_\mathrm{s}^{2} \,,
\end{equation}
such that $v_\mathrm{R} \approx 2\,{c_\mathrm{s}^\mathrm{in}}$ and $v_\mathrm{D} \approx \frac{c_\mathrm{s}^{2}}{2\,{c_\mathrm{s}^\mathrm{in}}} \ll 1\, \mathrm{km\,s^{-1}}$, with $c_\mathrm{s}^\mathrm{in} \approx \mathcal{O}(10 \,\mathrm{km\,s^{-1}})$ and $c_\mathrm{s} \approx \mathcal{O}(1 \,\mathrm{km\,s^{-1}})$.
The baryonic density within the ionized region can be written as:
\begin{equation} \label{eq:rhoin} 
 \rho^\mathrm{in}  = 
\begin{dcases}
       \rho^\mathrm{in}_{-} &   \quad\text{for } v_\mathrm{pbh}\geq v_\mathrm{R}\\
       \rho^\mathrm{in}_{0} &   \quad\text{for } v_\mathrm{D} < v_\mathrm{pbh} < v_\mathrm{R}\\ 
       \rho^\mathrm{in}_{+} &   \quad\text{for } v_\mathrm{pbh}\leq v_\mathrm{D}\,,
\end{dcases}\,,
\end{equation}
with
\begin{equation} 
\label{eq:rhoinpre}
 \rho^\mathrm{in}_{0} = \rho \left(\frac{v_\mathrm{pbh}^{2} + c_\mathrm{s}^{2}}{2 \,{c_\mathrm{s}^\mathrm{in}}^{2}}\right)\,, \quad\text{and}\quad \rho^\mathrm{in}_{\pm} = \rho \,\left(\frac{v_\mathrm{pbh}^{2} + c_\mathrm{s}^{2} \pm \sqrt{\Delta}}{2 \,{c_\mathrm{s}^\mathrm{in}}^{2}}\right)\,.
\end{equation}

\begin{figure}[tb!]
\centering
\includegraphics[width=0.6\textwidth]{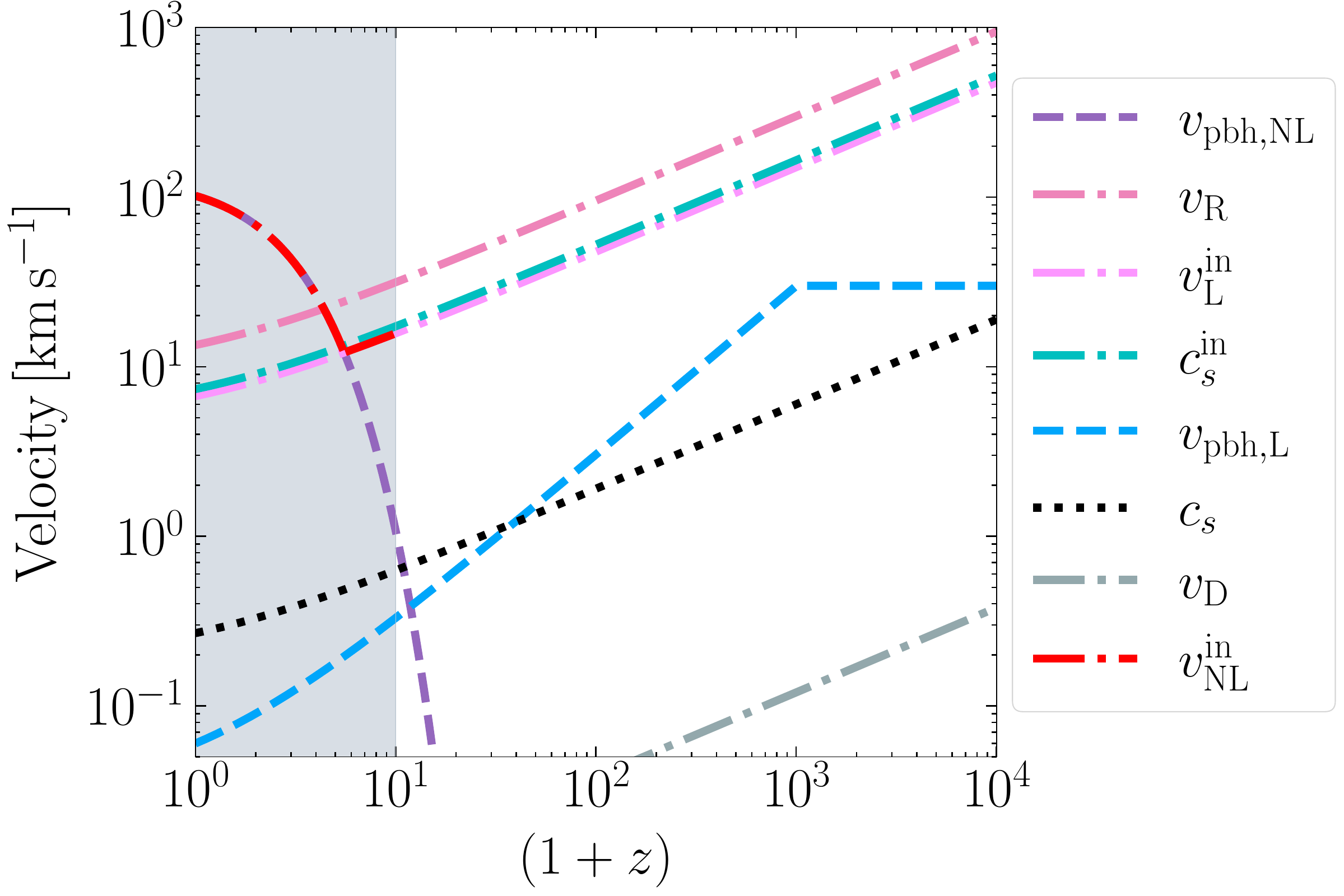}
\caption{Variation of the speed of sound and velocity of PBHs inside and outside the ionized region in the PR accretion model as a function of redshift. Here, the values of $c_{s}$ and $v_\mathrm{pbh}$ in the non-ionized region are selected as per Ref.~\cite{Serpico:2020ehh}, given by Eqs.~\eqref{eq:c_serpico} and~\eqref{eq:vpbhfull} respectively. The superscript ``$\mathrm{in}$'' represents the velocity profiles influenced by the radiative feedback within the ionized region, as signified by Eq.~\eqref{eq:vin}. The subscripts ``L'' and ``NL'' refer to the linear and non-linear regimes respectively. The curves for $v_\mathrm{L}^\mathrm{in}$ and $c_s^\mathrm{in}$ are slightly offset from each other to improve visibility.} 
\label{fig:velPR}
\end{figure}

The variation of the velocity profiles inside and outside the ionized region in the PR model is depicted in Figure~\ref{fig:velPR}. The dot-dashed lines correspond to the values of $c_s$ and $v_\mathrm{pbh}$ within the ionized region, while the remaining lines represent the corresponding values outside it. 
In the linear regime, for the values of $c_s$ and $v_\mathrm{pbh}$ selected as per SPIK20, the velocity of PBHs lies within the range $v_\mathrm{D} < v_\mathrm{pbh} < v_\mathrm{R}$. Hence, we can write the PBH velocity inside the ionized region as $v^\mathrm{in} \equiv c_\mathrm{s}^\mathrm{in} = 25 \cdot c_s$ (Eq.~\eqref{eq:vin})
with the density of baryons inside the ionized region as $\rho^\mathrm{in} = \rho^\mathrm{in}_{0}$ (Eq.~\eqref{eq:rhoin}) such that the accretion rate becomes~\cite{Scarcella:2020ssk}:
\begin{equation} \label{eq:30}
\dot  M_{\mathrm{PR},0} = \pi \frac{\left(G M\right)^{2} }{\sqrt{2}} \, \rho\, \frac{\left(v_\mathrm{pbh}^{2} + c_{s}^{2}\right)}{{\left(c_\mathrm{s}^\mathrm{in}\right)}^{5}}\,.
\end{equation}
This expression denotes the accretion rate in the linear regime, taking into account radiative feedback as the key feature of the PR model. In the non-linear regime, the PBH velocity inside the ionized region increases significantly, as depicted by the red dot-dashed line in Fig.~\ref{fig:velPR}, with the potential to substantially reduce the accretion rate.

\begin{figure}[tb!]
\centering
\includegraphics[width = 0.65\textwidth]{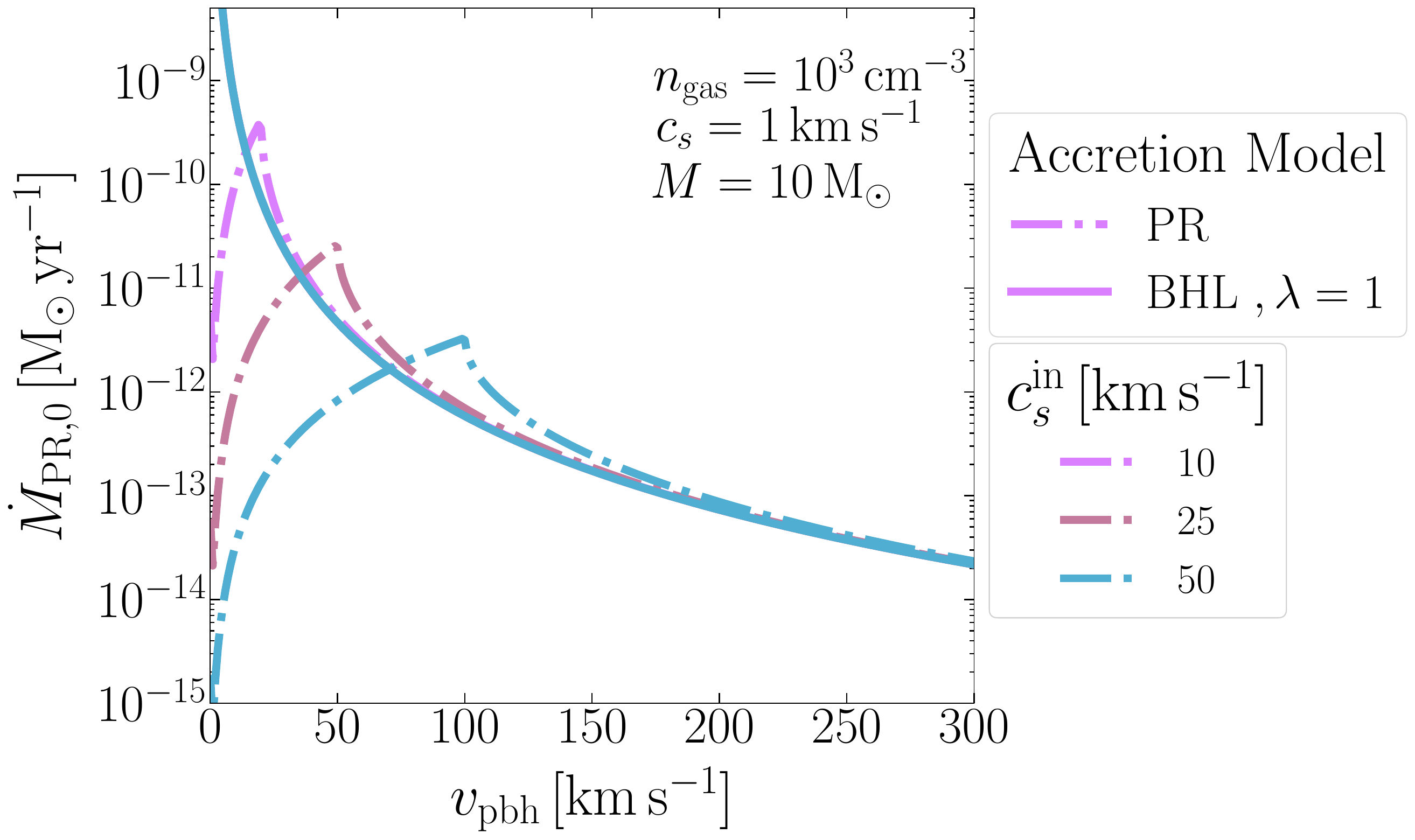}
\caption{Variation of the baryonic accretion rates of isolated PBHs in PR model, as a function of their velocity. This figure is analogous to Figure~$1$ of Ref.~\cite{Scarcella:2020ssk}. Here, the mean number density $n_\mathrm{gas}$ (and hence the density $\rho_{b}$ of the surrounding gas) and the speed of sound $c_\mathrm{s}$ are kept fixed, while the accretion efficiency is set to unity.}
\label{fig:MdotPR}
\end{figure}

Analogous to Ref.~\cite{Scarcella:2020ssk}, the dependence of the accretion rate $\dot{M}_\mathrm{PR}$ on the velocity of PBHs velocity is illustrated in Figure~\ref{fig:MdotPR}. This figure shows that unlike the BHL model, the accretion rates in PR model increase with increase in $v_\mathrm{pbh}$, reaching to maximum values at $v_\mathrm{pbh} = v_\mathrm{R} = 2\, c_\mathrm{s}^\mathrm{in}$. But, for $v_\mathrm{pbh}$ values larger than $2\, c_\mathrm{s}^\mathrm{in}$, the model predicts a further reduction of the accretion rate, approximating the dynamics predicted in the BHL model. So, Figure~\ref{fig:MdotPR} clearly illustrates how crucially the baryonic accretion rates in PR model rely on choice of the speed of sound and the velocity of the PBHs inside the ionized region.

\subsection{Accretion rates}
\label{sec:PR_accretion_rates}
As discussed earlier in Sec.~\ref{sec:Accretion around PBHs with DM halos}, the presence of DM halos around the PBHs enhances the gravitational potential, making accretion more efficient. For the PR model, the effective Bondi radius inside the ionized region $r_{\mathrm{B},\mathrm{eff}}^\mathrm{in}$ can be evaluated in the same way as described in Sec.~\ref{sec:Accretion around PBHs with DM halos}, making the identifications $v_\mathrm{eff}\rightarrow v_\mathrm{eff}^\mathrm{in}$ and $r_{\mathrm{B},h}\rightarrow r_{\mathrm{B},h}^\mathrm{in}$. We assume that the presence of DM halos does not affect the density and velocity profiles inside and outside the ionized region, provided the size of the DM halos is sufficiently smaller than the size of the ionized region.

\begin{figure}[tb!]
\centering
\includegraphics[width = 0.393\textwidth]{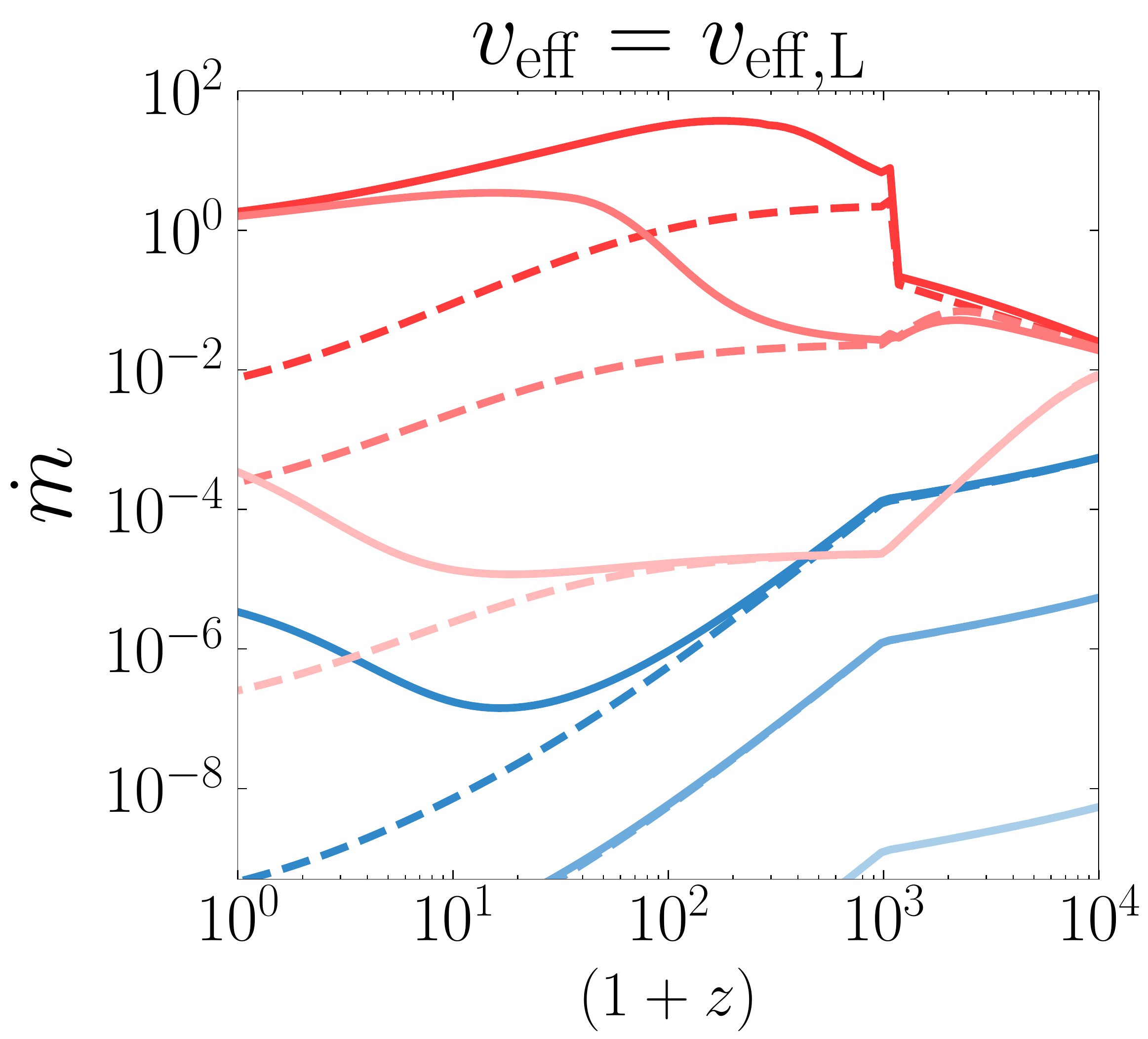}
\hfill
\includegraphics[width = 0.393\textwidth]{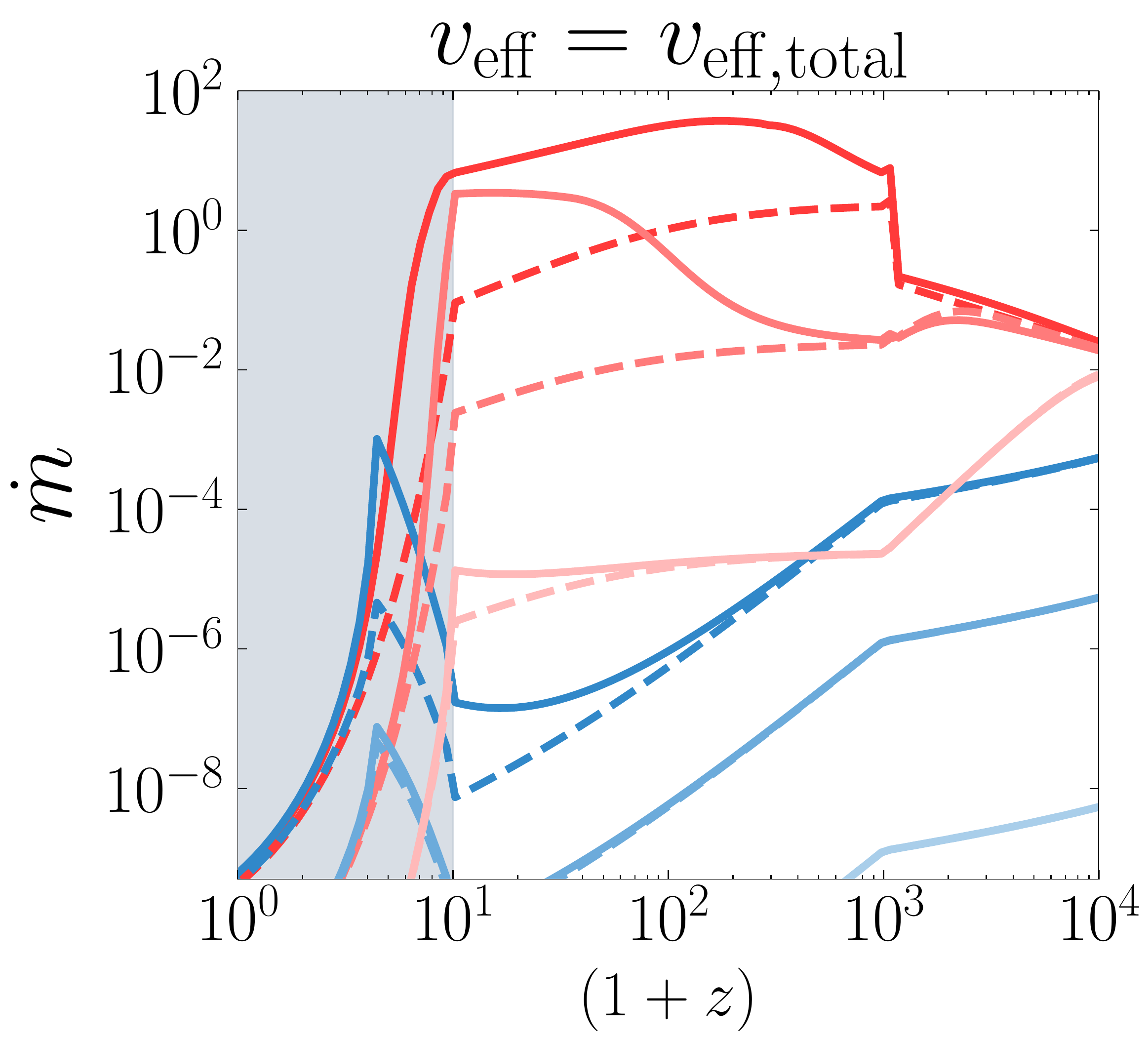}
\includegraphics[width = 0.19\textwidth]{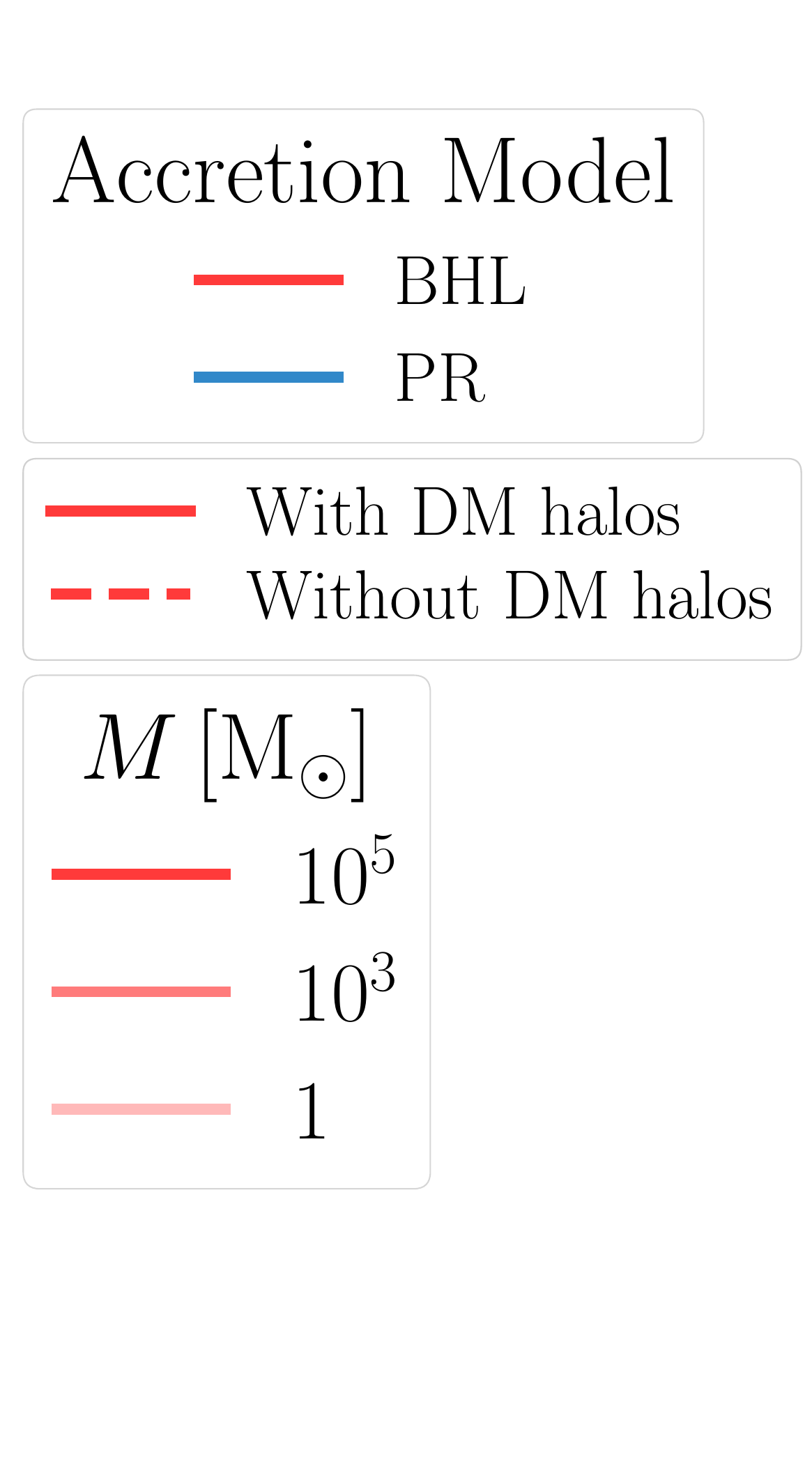}
\caption{Variation of the dimensionless accretion rate $\dot m$ as a function of the redshift $(1+z)$, in BHL and PR accretion models for PBHs with and without DM halos. The left panel illustrates the scenarios of PBHs evolving under the Hubble expansion only while the right panel also accounts for PBHs falling into virialized halos during structure formation in the late Universe (as shown by the
grey shaded region).}
\label{fig:mdotBHLPRcomp}
\end{figure}

\begin{figure}[tb!]
\centering
\includegraphics[width = 0.49\textwidth]{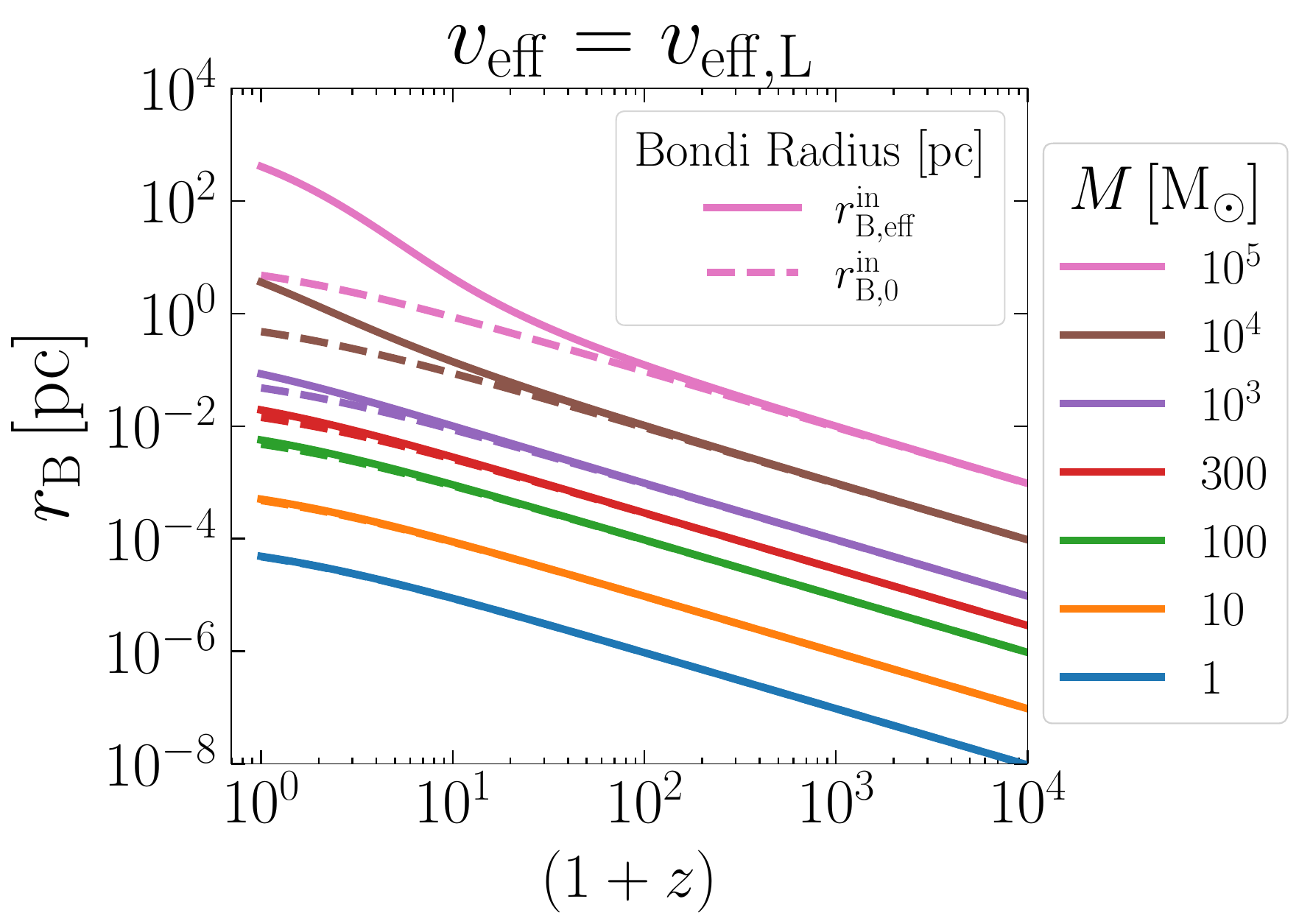}
\hspace*{\fill}
\includegraphics[width = 0.49\textwidth]{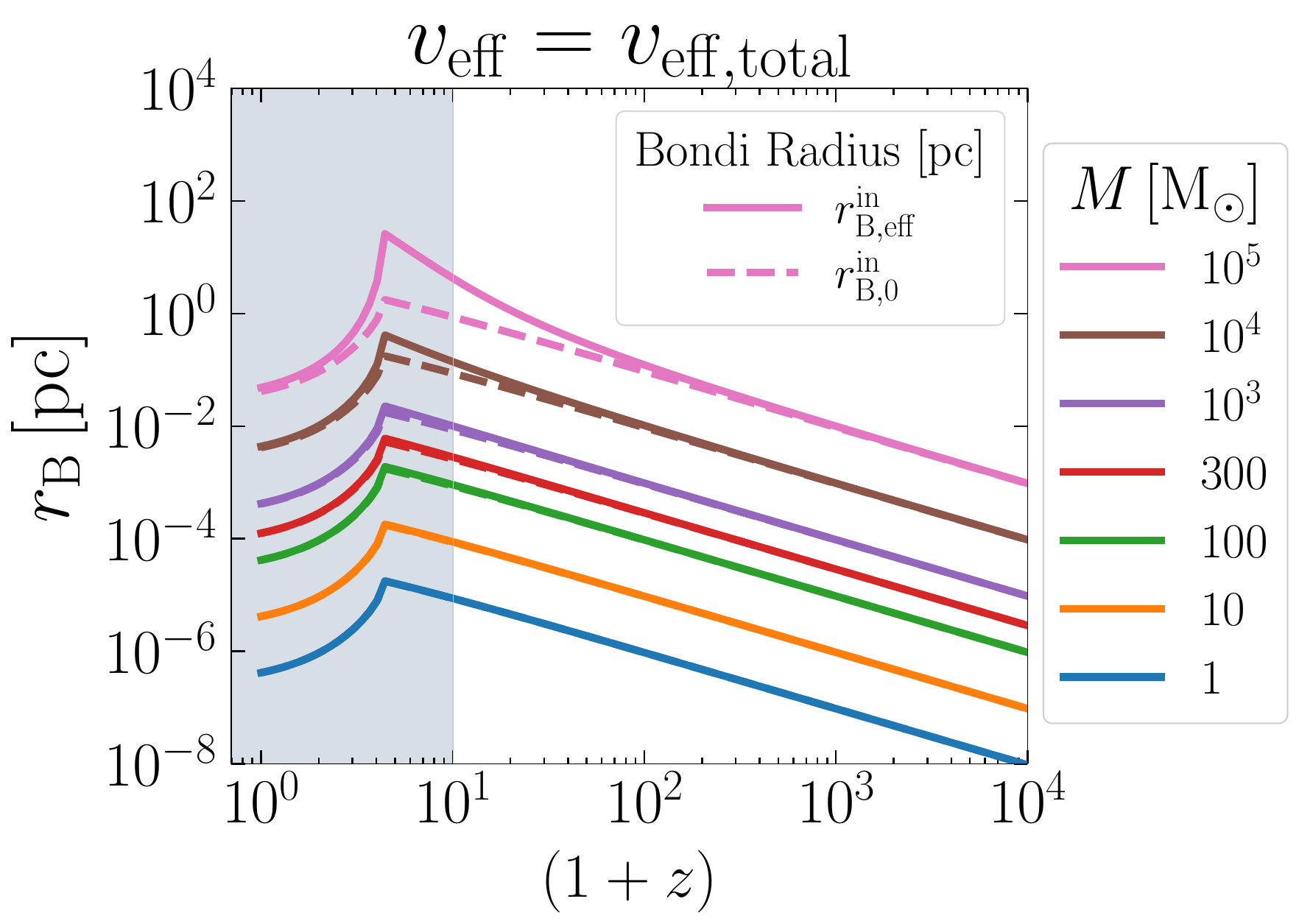}
\caption{Variation of Bondi radius $r_\mathrm{B}$ for PBHs with and without the presence of DM halos around the PBHs accreting as per PR model. The left panel shows the values of $r_\mathrm{B}$ for PBHs moving under the global cosmic expansion while the left panel depicts the corresponding values by including the effects of the fall of PBHs into virialized
halos for $z\leq 10$. }
\label{fig:BondiradiusPR}
\end{figure}

In the presence of DM halos, then, the accretion rate of PBHs in PR model becomes:
\begin{equation} \label{eq:MdotPRh}
  \dot M_{\mathrm{PR}, \, h}  = 4 \pi \, \rho_\mathrm{in} \, \, v_\mathrm{eff}^\mathrm{in}\, {r_\mathrm{B,eff}^\mathrm{in}}^{2}\,.
\end{equation}
The variation of the dimensionless accretion rates of PBHs with and without DM halos is shown in Figure~\ref{fig:mdotBHLPRcomp}. The red lines show the accretion rates for the BHL model (reproducing selected lines from Fig.~\ref{fig:mdotBHL}), while the blue lines show the corresponding accretion rates in the PR model. At high redshift, the accretion rates in the PR model are approximately $100$ times smaller than the corresponding values in the BHL model. This gap increases drastically with decrease in redshift after $z=1000$. This behavior can be explained by the fact that $\dot{m}_\mathrm{PR}$ is directly proportional to the baryonic density inside the ionized region $\rho^\mathrm{in}$. The decrease in the PBH velocity and sound speed after $z < 1000$ leads to a significant reduction in $\rho^\mathrm{in}$ (see Eq.~\eqref{eq:rhoin}). In contrast to the BHL model, in the non-linear regime, accretion rates in the PR model initially increase below $z = 10$, reaching a peak value and then decreasing, in line with the BHL model. This is due to the rapid increase in $v_\mathrm{pbh}$ and $v^\mathrm{in}$, leading to a rise in the Bondi radius inside the ionized region (Figure~\ref{fig:BondiradiusPR}) and hence in $\dot{m}_\mathrm{PR}$ until $v_\mathrm{pbh} = v_\mathrm{R}$. Beyond this, as $v_\mathrm{pbh}$ becomes larger than $v_\mathrm{R}$, $\dot{m}_\mathrm{PR}$ declines sharply due to the faster motion of the PBHs, transitioning to the supersonic regime similar to the BHL model (shown in Figure~\ref{fig:MdotPR}). We also see from Fig.~\ref{fig:MdotPR} that increasing the sound speed in the ionized bubble further suppresses the accretion rate. Indeed, we find that varying $c_s^\mathrm{in}$ in the range $10 - 50\,\mathrm{km/s}$ (for $c_s = 1 \,\mathrm{km/s}$) leads to a variation in the accreted mass of around 5 orders of magnitude, further highlighting the uncertainties associated with radiative feedback.

\section{Accreted masses of PBHs}
\label{subsec:Accreted masses of PBHs and their Velocity dependence}
\begin{figure}[tb!]
\centering
\includegraphics[width = 0.49\textwidth]{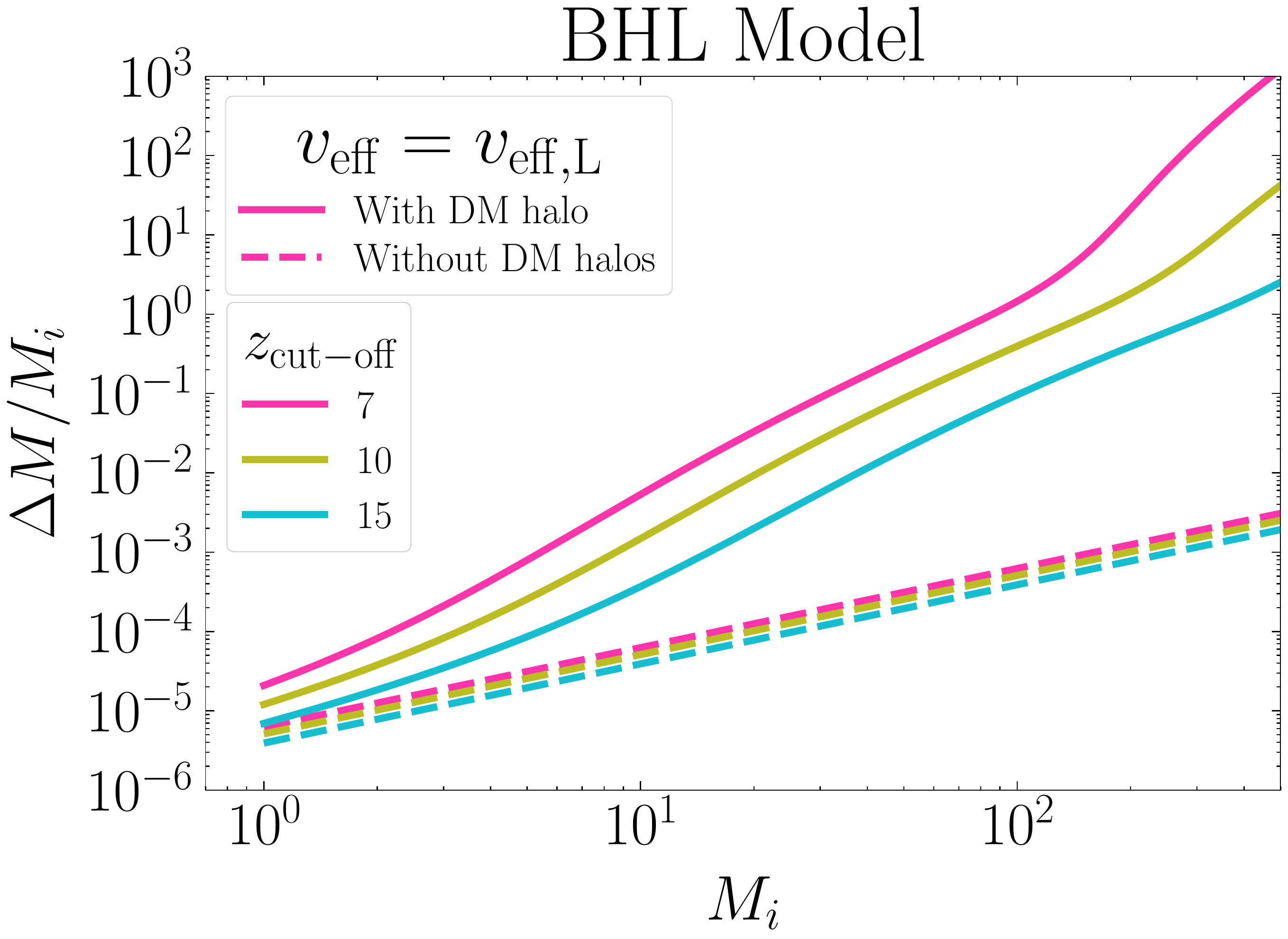}
\hspace*{\fill}
\includegraphics[width = 0.49\textwidth]{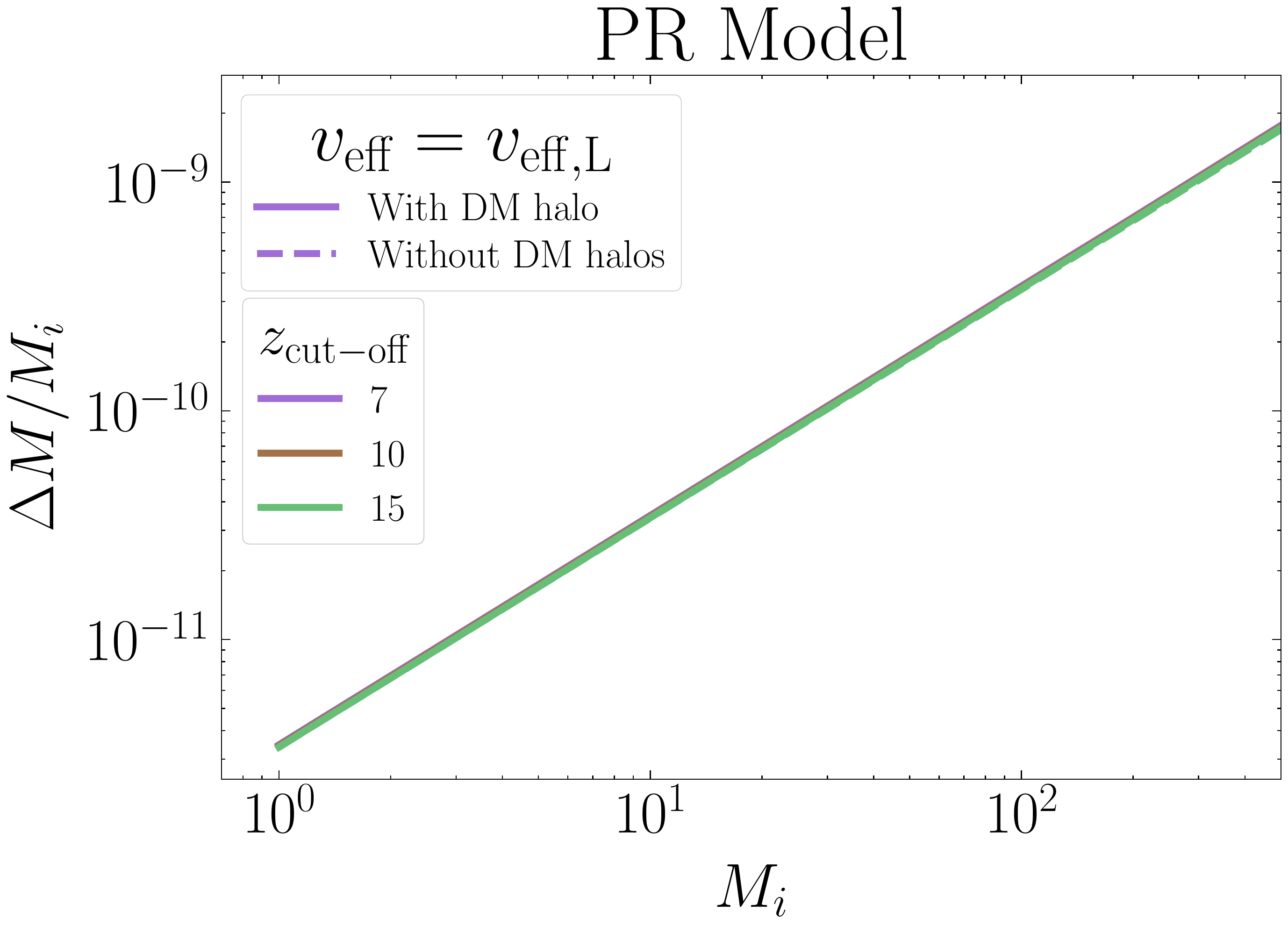}
\caption{Fractional increase in the mass of the PBHs in BHL and PR accretion models, as a function of their initial mass $M_i$. The final PBH mass is $M_f(z = z_\mathrm{cut-off})$, leading to $\Delta M = (M_f(z = z_\mathrm{cut-off}) - M_i)$ for different values of $z_\mathrm{cut-off}$. The dashed lines represent PBHs without DM halos and the solid lines designate PBHs with DM halos. Similar to ROM07, the constant electron fraction after recombination is assumed to be $x_e = 10^{-3}$~\cite{Ricotti:2007au}. In the right panel, the scenarios with and without DM halos are indistinguishable. Only baryonic accretion is included; an initial period of radiation accretion should also be accounted for, as described in Sec.~\ref{sec:Accretion of radiation around PBHs}.}
\label{fig:MfBHLPRSPIK20}
\end{figure}

The results of the previous section highlight that the PR model features comparatively lower accretion rates than the BHL model, emphasizing the importance of feedback in shaping accretion rates. In this section, we provide a comparison of the total accreted mass in both models, as well as exploring the uncertainties arising from the choices of velocity profiles.  We evaluate the final PBH mass $M_f(z = z_\mathrm{cut-off} )$ with $z_\mathrm{cut-off} = 15, \,10,\, 7$, by integrating the baryonic accretion rates shown in Figures~\ref{fig:mdotBHL} and~\ref{fig:mdotBHLPRcomp}.
The fractional increase in PBH mass $\Delta M(z_\mathrm{cut-off})/M_i$ as a function of $M_i$ is illustrated in Figure~\ref{fig:MfBHLPRSPIK20} for accretion in the BHL model (left) and PR model (right). 

Both the panels of this figure show that the fractional change in the PBH mass increases with increase in their initial mass $M_i$. The left panel of this figure illustrates that in the BHL model, the growth in the PBH mass is larger for PBHs with DM halos. This results from the relatively higher accretion rates of PBHs with DM halos shown in Figure~\ref{fig:mdotBHL}. We also find that in the BHL model, for PBHs with initial mass $M_i \geq 100\, \mathrm{M_{\odot}}$, baryonic accretion can increase their masses by several orders of magnitude when accounting for the presence of DM halos. Moreover, we verified that in the non-linear regime, the growth in PBH masses is suppressed due to the sharp decline in accretion rates (as illustrated in Figure~\ref{fig:mdotBHL}). We also checked that these outcomes are the same whether the electron fraction $x_e$ is either $10^{-3}$ or $1$, post-recombination. This is due to the fact that the contribution of the electron fraction $x_e$ terms in the accretion efficiency $\lambda$ becomes less and less efficient at late times. Given the uncertainties associated with the value of the accretion efficiency $\lambda$, we have also considered an alternative approach, fixing the value to $\lambda = 0.1$ for all redshift. We find that in this case the change in PBH mass is always less than $60\%$ over the range of initial masses shown in the left panel of Fig.~\ref{fig:MfBHLPRSPIK20}. This indicates that the large growth of mass in BHL model relies on the saturation of $\lambda \sim 1$ which occurs after recombination (see Fig.~\ref{fig:lambda}).

The right panel of Fig.~\ref{fig:MfBHLPRSPIK20} shows that, in comparison to the BHL model, the fractional change in the masses of PBHs with and without DM halos in the PR model is completely negligible. This difference results from the relatively higher values of the speed of sound and velocity of PBHs inside the ionized region leading to smaller Bondi radii in the PR model. We also note that the final PBH mass does not depend on the cut-off redshift, because in the PR model, the accretion rate drops rapidly at late times (as illustrated in Fig.~\ref{fig:mdotBHLPRcomp}). As the PBH velocity decreases at low redshift, the effects of feedback become more prominent and the accretion rate drops (see Fig.~\ref{fig:MdotPR}). These results imply that in the PR model, the growth of PBH masses by baryonic accretion is negligible, compared to the substantial growth predicted by the BHL model for large PBH masses. 

We therefore find that a significant increase in PBH masses can only be obtained invoking a specific set of assumptions: accretion according to the BHL model, enhanced by the presence of DM halos, and high accretion efficiency (with $\lambda \sim 1$ during some period of cosmic history).

\subsection{Velocity dependence of $\Delta M/M_i$}
\label{subsec:Velocity dependence}
In general, the efficiency of accretion around PBHs and hence the change in mass of the PBHs via accretion crucially depends on various accretion parameters such as gas viscosity, Hubble expansion, the speed of sound and the PBH velocity. To explicitly quantify the dependence of the PBHs accretion rates on the velocity profiles, we select the values of $c_s$ and $v_\mathrm{pbh}$ as per ROM07~\cite{Ricotti:2007au} and compare the accretion scenarios discussed earlier using the velocity profiles of SPIK20~\cite{Serpico:2020ehh} (shown in Figure~\ref{fig:velSPIK20}). Full details of the accretion rates calculated using the velocity profiles of ROM07 are discussed in Appendix~\ref{sec:Sensitivity of the accretion rate on the choice of velocity profiles}, with the ROM07 velocity profiles shown in Fig.~\ref{fig:velROM07BHLPR}.

\begin{figure}[tb!]
\centering
\includegraphics[width=0.49\textwidth]{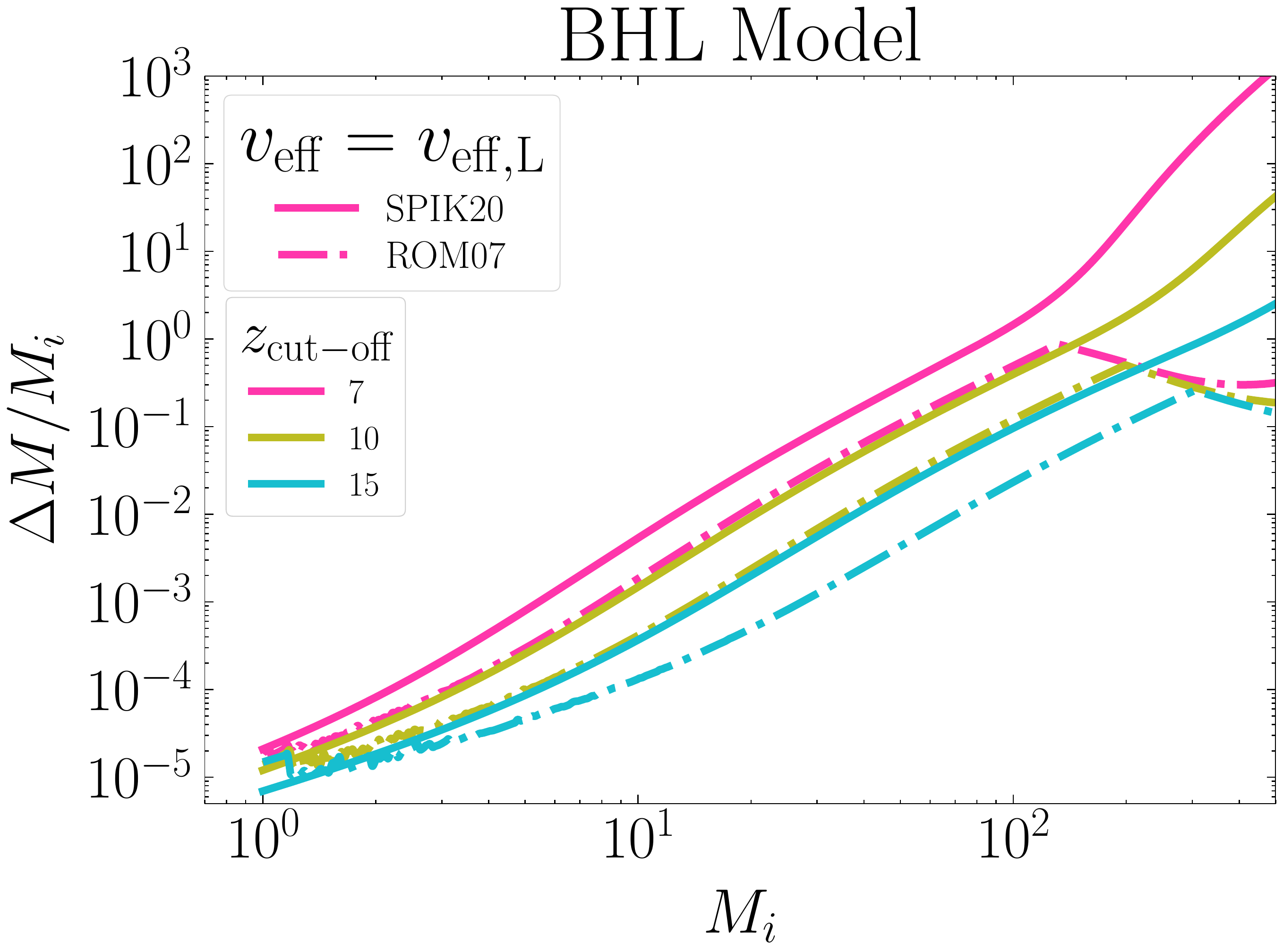}
\hspace*{\fill}
\includegraphics[width=0.49\textwidth]{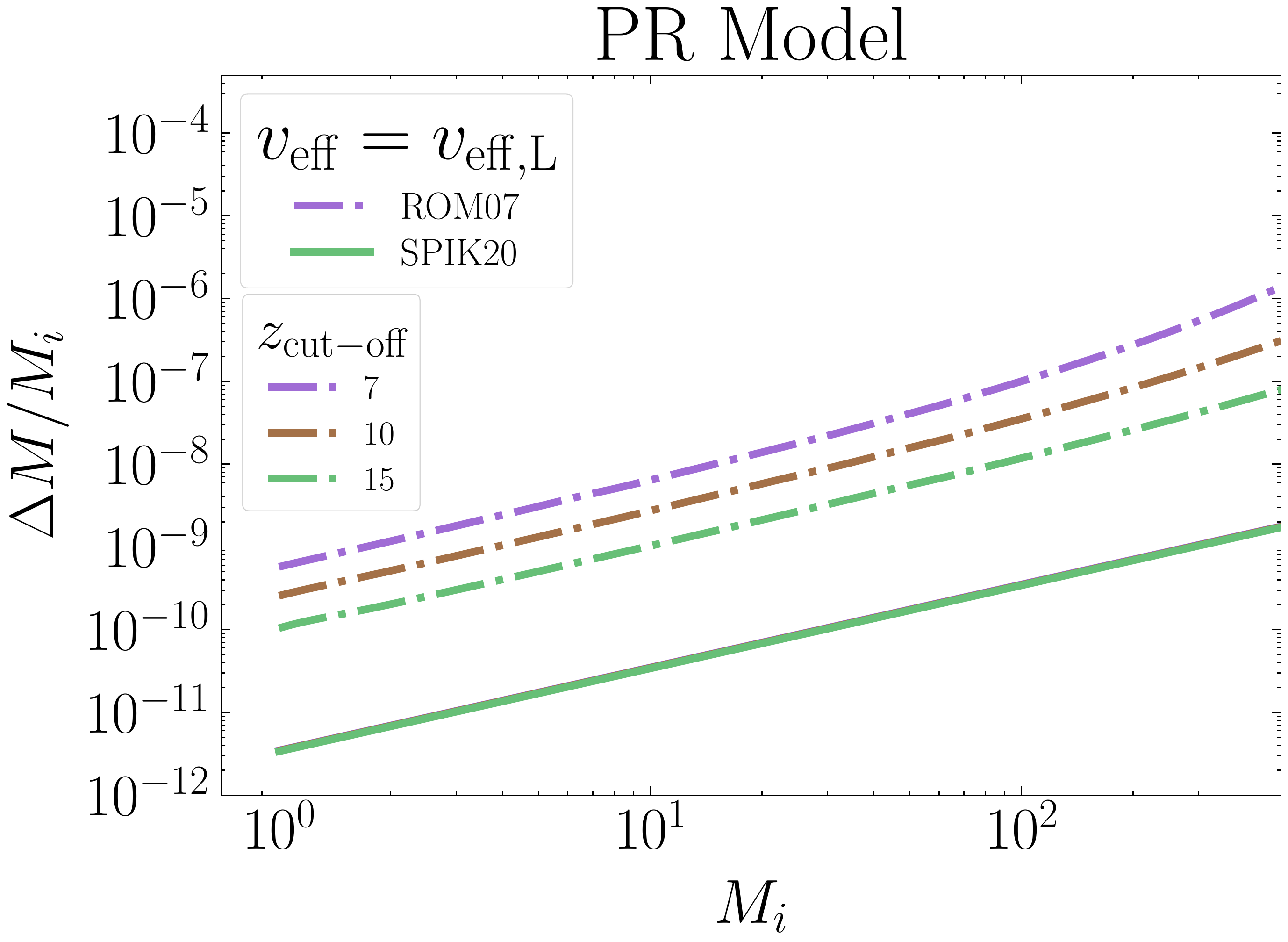}
\caption{Fractional increase in the mass of the PBHs with DM halos, as a function of their initial mass. The left panel shows the scenario for BHL model while the right panel illustrates the case in PR model, with $v_\mathrm{eff} = v_\mathrm{eff,L}$ estimated as per velocity profiles of SPIK20 and ROM07.}
\label{fig:MfBHLPRROM07}
\end{figure}

The fractional change in PBH mass in the linear regime for PBHs with DM halos in the BHL and PR models (evaluated as per the velocity profiles of SPIK20 and ROM07) is shown in Figure~\ref{fig:MfBHLPRROM07}. The left panel of this figure shows that in the BHL model, the final mass of PBHs at $z_f = z_\mathrm{cut-off}$ can vary significantly depending on whether $c_s$ and $v_\mathrm{pbh}$ are based on ROM07 or SPIK20. The velocity profiles as per ROM07 typically reduce the final accreted mass by a factor of approximately $10$ compared to SPIK20, for initial PBH masses $M_i \sim M_\odot$. At higher masses, $M_i \sim 100 \,M_\odot$, the ROM07 velocity profiles reduce the accreted mass by as much as 5-6 orders of magnitude. This difference occurs because ROM07 predicts a larger PBH-gas relative velocity, leading to lower accretion rates (see Figure~\ref{fig:mdotcompBHL}), especially for $z \leq 100$.

Furthermore, the right panel of Fig.~\ref{fig:MfBHLPRROM07} shows that in the PR model, the final accreted mass is enhanced by 4-5 orders of magnitude when the speed of sound and PBH velocity follow ROM07 rather than SPIK20. This disparity arises due to the smaller sound speed of the background gas in the ROM07 model. This leads to a smaller sound speed in the ionized bubble and thus a larger gas density. Though the growth in PBH mass remains negligible in the PR model (at least for $M_i \lesssim 100\, M_\odot$, see Figure~\ref{fig:mdotcompPR}), the large difference in results emphasizes the delicate dependence of the baryonic accretion rates on assumptions about the velocity profiles.


\section{Conclusion}
\label{sec:Conclusion}
In this paper, we have explored the relevance of accretion for PBHs revisiting the Bondi-Hoyle-Lyttleton (BHL) and Park-Ricotti (PR) models. We studied the accretion rates of radiation around isolated PBHs, up to matter-radiation equality. Then, we extended our analysis to baryonic accretion for PBHs with and without DM halos, in the linear and non-linear regimes, corresponding to the homogeneous Hubble expansion and structure formation in the late Universe, respectively. 
We also analyzed the impact of key accretion parameters such as the speed of sound and velocity of the PBHs selected as per Serpico et al.~\cite{Serpico:2020ehh} (SPIK20) and Ricotti et al.~\cite{Ricotti:2007au} (ROM07).

Firstly, the accretion of radiation in the early Universe can increase the masses of the PBHs up to $4 \%$ (for accretion efficiency $\lambda = 0.1$), irrespective of their initial masses. The accretion of radiation is dominant only at very early times, due to the rapid dilution of the radiation density with redshift. While more massive PBHs can in principle have a larger accretion rate due to their larger mass, they also form at later times, when the radiation density is lower. This results in a saturation value for the accretion of radiation, independent of the masses of the PBHs. We also point out that accurately modeling the time of PBH formation is crucial for an accurate estimate of the growth in PBH due to radiation accretion. The final PBH mass is also sensitive to the accretion efficiency $\lambda$ (saturating at a $60\%$ increase for $\lambda = 1$) and we highlight that the uncertainty in this value in the early Universe may prevent precise predictions.

Regarding baryonic accretion, we find that for isolated PBHs in the linear regime, the PR model generally predicts significantly lower accretion rates compared to the BHL model. More specifically, for PBHs with mass $\geq 10^{3} \, \mathrm{M_{\odot}}$, the accreted mass in the PR model can be approximately $\mathcal{O}(10^{7})$ times smaller than the corresponding values in the BHL model. This notable reduction in the PR model arises due to the radiative feedback leading to the formation of an ionization front (as shown in Figure~\ref{fig:PRmodel}) and the elevation of the speed of sound inside the ionized region. At lower redshift, the prior existence of DM halos around massive PBHs can narrow down the gap between the accretion rates of the BHL and PR models (as depicted in Figure~\ref{fig:mdotBHLPRcomp}). In the non-linear regime, we observe a sharp decrease in accretion rates for $z \leq 10$, resulting from  the rapid increase in the velocity of PBHs during structure formation. This abrupt enhancement in the PBHs velocity is so strong that both in BHL and PR models, the accretion rates get suppressed by the same order resulting in equivalent values in both cases. 

We could also consider an alternative scenario in which a PBH settles to the centre of a DM halo. Such a scenario might be proposed to explain the rapid growth of Supermassive Black Holes (SMBHs) from intermediate mass seeds, with the low PBH velocity with respect to the surrounding gas giving an enhancement of the accretion rate. However, as seen in Fig.~\ref{fig:velPR}, the effective PBH velocity is dominated by the sound speed at $z \lesssim 30$, so such a scenario cannot substantially enhance the accretion rate over that observed in the linear regime (in which the PBH velocity effectively vanishes at late times).

In addition, we have compared the accretion rates of PBHs using two types of velocity profiles designating the speed of sound $c_s$ and the velocity  $v_\mathrm{pbh}$ of the accreting PBHs. Selecting these values as per ROM07 and SPIK20 leads to vastly different accretion rates. For massive PBHs having DM halos, the choice of $c_s$ and $v_\mathrm{pbh}$ as per ROM07 even predicts higher accretion rates for PR model, defying our earlier observations. Furthermore, for velocity profiles selected as per SPIK20, the BHL model predicts that PBHs with initial masses larger than $100\, \mathrm{M_{\odot}}$ can grow by many orders of magnitude by low redshifts ($z \lesssim 15$). However, the fractional change in the PBH mass assuming velocities from ROM07 is approximately 5 orders of magnitude smaller than in the SPIK20 scenario. On the contrary, regardless of the choice of velocity parameters, the PR model predicts a negligible change in the masses of PBHs due to the accretion of baryons. 

Figure~\ref{fig:MfBHLPRROM07} clearly shows that apart from in the BHL model in presence of DM halos, the increase in PBH masses due to baryonic accretion is negligible, regardless of the detailed modeling and masses of the PBHs. A large increase in PBH masses could potentially affect PBH mass functions and constraints, as well as being relevant for the growth of early SMBHs. However, our results imply that such a large mass increase can be obtained only under a specific set of assumptions: when combining the BHL accretion model, the specific formalism for the accretion efficiency $\lambda$ described in ROM07, and the accretion boost due to the presence of DM halos around PBHs. When one of these assumptions is relaxed, for instance in the presence of radiation feedback as in the PR model, the change in the PBH mass over cosmic history is almost negligible in the whole mass range under consideration, regardless of the details of the models themselves and the redshift down to which accretion is considered. These results are also sensitive to some key parameters such as the speed of sound and velocity of PBHs. However, we find that different choices for the modeling of these parameters again can lead to a drastic reduction in the accreted mass.

Given these considerations, it is by no means guaranteed that PBHs undergo substantial growth by accretion of either radiation or baryons. So, we emphasize the need for further study of the interplay between radiative feedback, the presence of DM halos, the accretion efficiency, and the PBH velocity profiles in order to provide a more robust assessment of the increase in PBH masses over cosmic time.


\begin{acknowledgments}
The authors thank V. De Luca for providing important insights regarding their work published in Ref.~\cite{DeLuca:2020fpg}. PJ would also like to thank Dr. Daniele Gaggero for hosting her research stay at INFN, Pisa, Italy, funded under the CSIC-iMOVE23 (IMOVE23009) programme. The thorough expertise provided by DG were crucial for the development of the work presented in this paper. The support of the administration and Theoretical Physics Group at INFN, Pisa, also significantly enhanced the progress of our research. The authors also thank the members of the `Dark Collaboration at IFCA' working group for their cooperation during the development of this work. 

PJ acknowledges support of project PGC2018-101814-B-100 (MCIU/AEI/MINECO/FEDER, UE) Ministerio de Ciencia, Investigaci\'on y Universidades. BJK acknowledges funding from the \textit{Ram\'on y Cajal} Grant RYC2021-034757-I and the \textit{Consolidaci\'on Investigadora} Project \textsc{DarkSpikesGW}, reference CNS2023-144071, financed by MCIN/AEI/10.13039/501100011033 and by the European Union ``NextGenerationEU"/PRTR.
J.M.D. acknowledges support from project PID2022-138896NB-C51 (MCIU/AEI/MINECO/FEDER, UE) Ministerio de Ciencia, Investigaci\'on y Universidades.
\end{acknowledgments}


\appendix
\section{Fraction of PBHs in virialized halos}
\label{sec:Fraction of PBHs in virialized halos}
\begin{figure}[tb!]
\centering
\includegraphics[width = 0.45\textwidth]{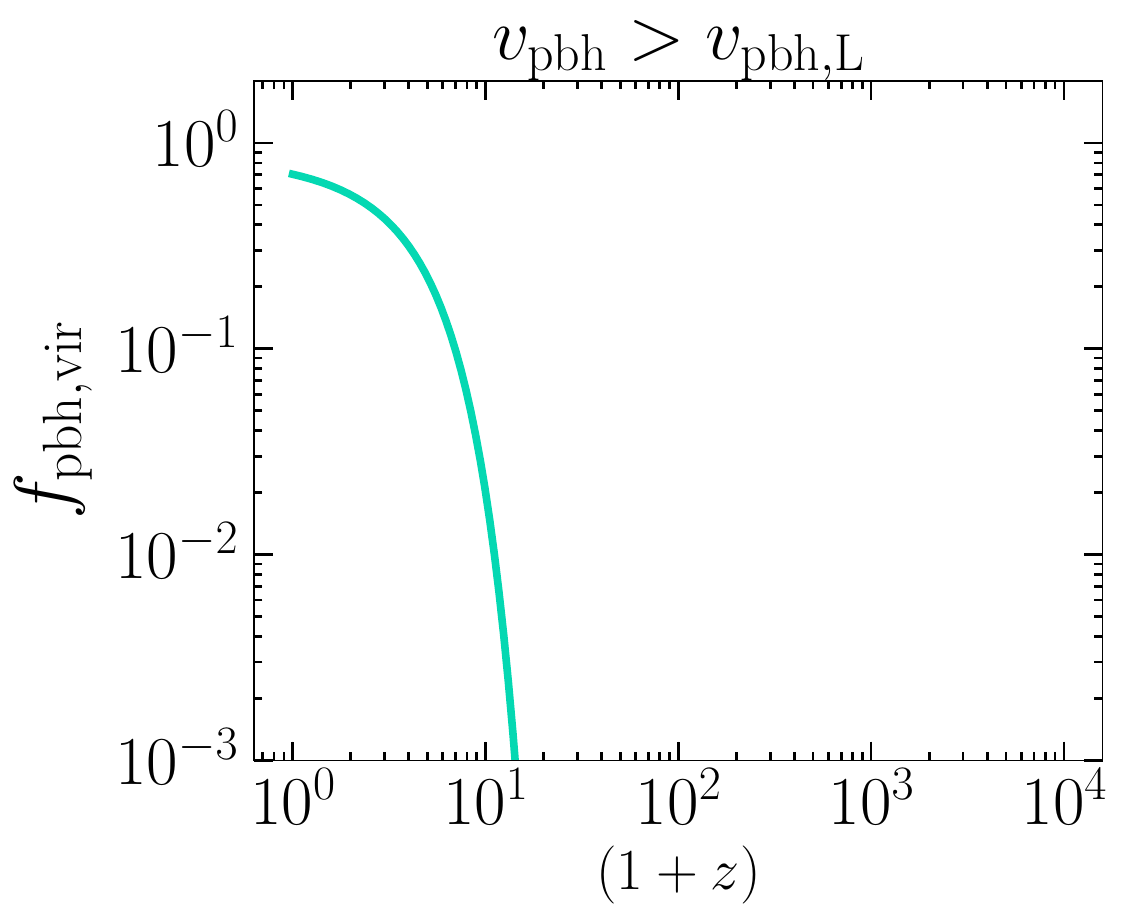}
\caption{Fraction of PBHs falling into virialized halos during structure formation in the late Universe, around $z\leq 10$. Here, we include PBHs falling into virialized halos with virial velocity $v > v_\mathrm{pbh,L}$ (given by Eq.~\eqref{eq:vpbhnonlinear}), at a given redshift $z$.}
\label{fig:fracvirPBH}
\end{figure}

In general, the process of structure formation can alter the velocity of PBHs. The fraction of PBHs falling into virialized structures with mass $M > M_\mathrm{min}$ can be written as~\cite{Ricotti:2007au}:
\begin{equation}  \label{eq:fpbhvir} 
f_\mathrm{pbh,vir}(z) = 1 - \mathrm{erf}\left(\frac{\delta_c}{\sqrt{2}\cdot\sigma(M_\mathrm{min}, z)}\right)\,.
\end{equation}
with 
\begin{equation}  \label{eq:sigmaMz} 
\sigma(M, z) \approx 10.2 - 0.79\,\mathrm{log}(M(z)) \,,
\end{equation}
being the variance of density perturbations of mass $M(z)$ given by Eq.~\eqref{eq:masssigma} and we assume an overdensity parameter $\delta_c =1.686$~\cite{Schaefer:2007nf}. 
Here, we provide an indication of the importance of structure formation by estimating the fraction of PBHs which are in virialized structures with virial velocity larger than the linear velocity $v_\mathrm{pbh,L}$ at that redshift. These PBHs will typically have their accretion rate suppressed compared to those which remain isolated. We thus fix $M_\mathrm{min}$ using the expression for the virial velocity $v_\mathrm{pbh,NL} \sim v_\mathrm{vir}$ in Eq.~\eqref{eq:vpbhnonlinear} and setting $v_\mathrm{pbh,NL} > v_\mathrm{pbh,L}$. The resulting fraction of PBHs is shown in Figure~\ref{fig:fracvirPBH}, suggesting that the effect of non-linear structure can become significant at redshifts below $z \lesssim 10$, with around $75\%$ of PBHs lying in structures with large virial velocities by $z = 0$.

\section{Sensitivity of the accretion rate on the choice of velocity profiles}
\label{sec:Sensitivity of the accretion rate on the choice of velocity profiles}
The phenomenon of accretion around PBHs can be influenced by many factors such as the speed of sound in the surrounding material, the velocity of the PBHs and the existence of DM halos around them prior to accretion. We have already discussed the impact of DM halos on the accretion rates of PBHs in both BHL and PR models in Sections~\ref{sec:Bondi-Hoyle-Lyttleton (BHL) Accretion model for baryons} and~\ref{sec:Park Ricotti (PR) Accretion model} respectively. So, in this Appendix we further explore how the values of the velocity profiles i.e.\ the speed of sound $c_s$ and proper velocity $v_\mathrm{pbh}$ of PBHs can modify their baryonic accretion rates. To do that, we select two sets of velocity profiles as per Ref.~\cite{Ricotti:2007au} (ROM07) and Ref.~\cite{Serpico:2020ehh} (SPIK20) and then compare the accretion rates of PBHs with and without DM halos in BHL and PR models. Through this comparative analysis, we aim to examine the uncertainties in the accretion process of PBHs which come with the choice of the speed of sound and velocity of the PBHs.

The velocity profiles selected as per SPIK20 are already detailed (Figure~\ref{fig:velSPIK20}) and discussed earlier in Section~\ref{sec:Bondi-Hoyle-Lyttleton (BHL) Accretion model for baryons}. As per ROM07, the speed of sound $c_s$ is given as:
\begin{equation}  \label{eq:A1} 
c_\mathrm{s} \simeq   5.7 \, \left(\frac{1 + z}{1000}\right)^{1/2} \,  \left[\left(\frac{1 +z_\mathrm{d}}{ 1 + z}\right)^{\beta} + 1\right]^{-1/\left(2 \, \beta\right)} \, \mathrm{km\,s^{-1}} \,,
\end{equation}
where $z_\mathrm{d} \approx 1100$ is the redshift at which the baryonic gas decouples from the photons and $\beta = 1.72$. Here, the speed of sound is calculated based on the gas temperature as a function of redshift. The value of $v_\mathrm{pbh}$ selected as per ROM07 is valid in the linear regime, evaluated with the  assumption that the PBHs move with the same velocity as that of the DM particles in the intergalactic medium. Hence, the velocity $v_\mathrm{pbh}$ of PBHs is equivalent to $\left\langle  v_\mathrm{rel} \right\rangle = \left\langle  v_\mathrm{DM} \right\rangle - \left\langle  v_\mathrm{gas} \right\rangle$, where $\left\langle  v_\mathrm{DM} \right\rangle$ and $\left\langle  v_\mathrm{gas} \right\rangle$ are the mean velocity of DM particles and baryons respectively, calculated in a small patch of the Universe from the corresponding power spectra (see ROM07, especially Fig.~2, for further details).

In practice, the PBH velocities will follow a Maxwellian distribution $f_M \left(v, \sigma\right)$ with width $\sigma \sim v_\mathrm{rel}$. We therefore take the effective velocity equal to the ensemble average, given by:
\begin{equation}
    \left\langle v_\mathrm{eff} \right\rangle = \int_{0}^{\infty} \left(c_{s}^{2}+v^{2}\right)^{1/2} f_M \left(v, \sigma\right)\, \mathrm{d}v = \frac{1}{\sqrt{2\pi}}\frac{c_s}{\mathcal{M}}e^{1/(4\mathcal{M})^2} K_1\left(\frac{1}{4 \mathcal{M}^2}\right)\,,
\end{equation}
where $K_{n}(x)$ denotes the modified Bessel function of the second kind and $\mathcal{M} = \left\langle  v_\mathrm{rel} \right\rangle/c_\mathrm{s}$ represents the cosmic Mach number. Since the accretion luminosity $L = \lambda \dot M c^{2}$ relies on the accretion rate of the PBHs, so this too depends on the velocity profiles and hence the effective velocities. As pointed out in ROM07~\cite{Ricotti:2007au}, depending on the values of the accretion rate $\dot m$, the accretion luminosity $L$ can characterized by two different power-law regimes of the effective velocity $v_\mathrm{eff}$, indicated as~\cite{Ricotti:2007au}:
\begin{equation} \label{eq:A3}
L \propto
\begin{dcases}
v_\mathrm{eff}^{-6} &,\: \text{if } \dot{m} < 1\,,\\
v_\mathrm{eff}^{-3} &,\:\text{if } \dot{m} \gtrsim 1\,.
\end{dcases}
\end{equation}
It is therefore useful to define the luminosity-weighted effective velocity $v_\mathrm{eff}$ as:
\begin{equation} \label{eq:A2}
\left\langle  v_\mathrm{eff}^{-\alpha} \right\rangle^{-1/\alpha}  = \left[\int_{0}^{\infty} \frac{f_M \left(v, \sigma\right)}{\left(c_{s}^{2}+v^{2}\right)^{\alpha/2}}\, \mathrm{d}v \right]^{-1/\alpha}\,.
\end{equation}
Now, for $\dot{m} < 1$ ($\alpha = 6$) we obtain:
\begin{equation} \label{eq:A4}
\frac{\left\langle  v_\mathrm{eff} \right\rangle_\mathrm{A}}{c_s} =
       \frac{2^{7/12} \mathcal{M}^{7/6}} { \left[\sqrt{2}\, \left(\mathcal{M} + \mathcal{M}^{3}\right) +  \sqrt{\pi}\, \left(-1 - 2\mathcal{M}^{2} + \mathcal{M}^{4}\right)\,e^{1/(2\mathcal{M}^{2})} \, \mathrm{erfc}(\frac{1}{\sqrt{2}\mathcal{M}})\, \right]^{1/6}}  \,,
\end{equation}
and for $\dot{m} > 1$ ($\alpha = 3$) we obtain:
\begin{equation} \label{eq:A5}
\frac{\left\langle  v_\mathrm{eff} \right\rangle_\mathrm{B}}{c_s} = 
       \frac{\sqrt{2}\,\pi^{1/6} \mathcal{M}^{5/3} e^{-1/(12\mathcal{M}^{2})}} { \left[ \left(1 + 2\,\mathcal{M}^{2}\right)  \cdot K_{0}\left( \frac{1}{4\mathcal{M}^{4}}\right)\, - \, \, K_{1}\left( \frac{1}{4\mathcal{M}^{4}}\right) \right]^{1/6}} \,.
\end{equation} 
While these expressions do not enter the mass accretion rate, they may be relevant for estimating the accretion luminosity. We report them here in full because we find a discrepancy with the expressions reported in ROM07, though we have verified the expressions above by numerical integration of Eq.~\eqref{eq:A2}.

\begin{figure}[tb!]
\centering
\includegraphics[width = 0.6\textwidth]{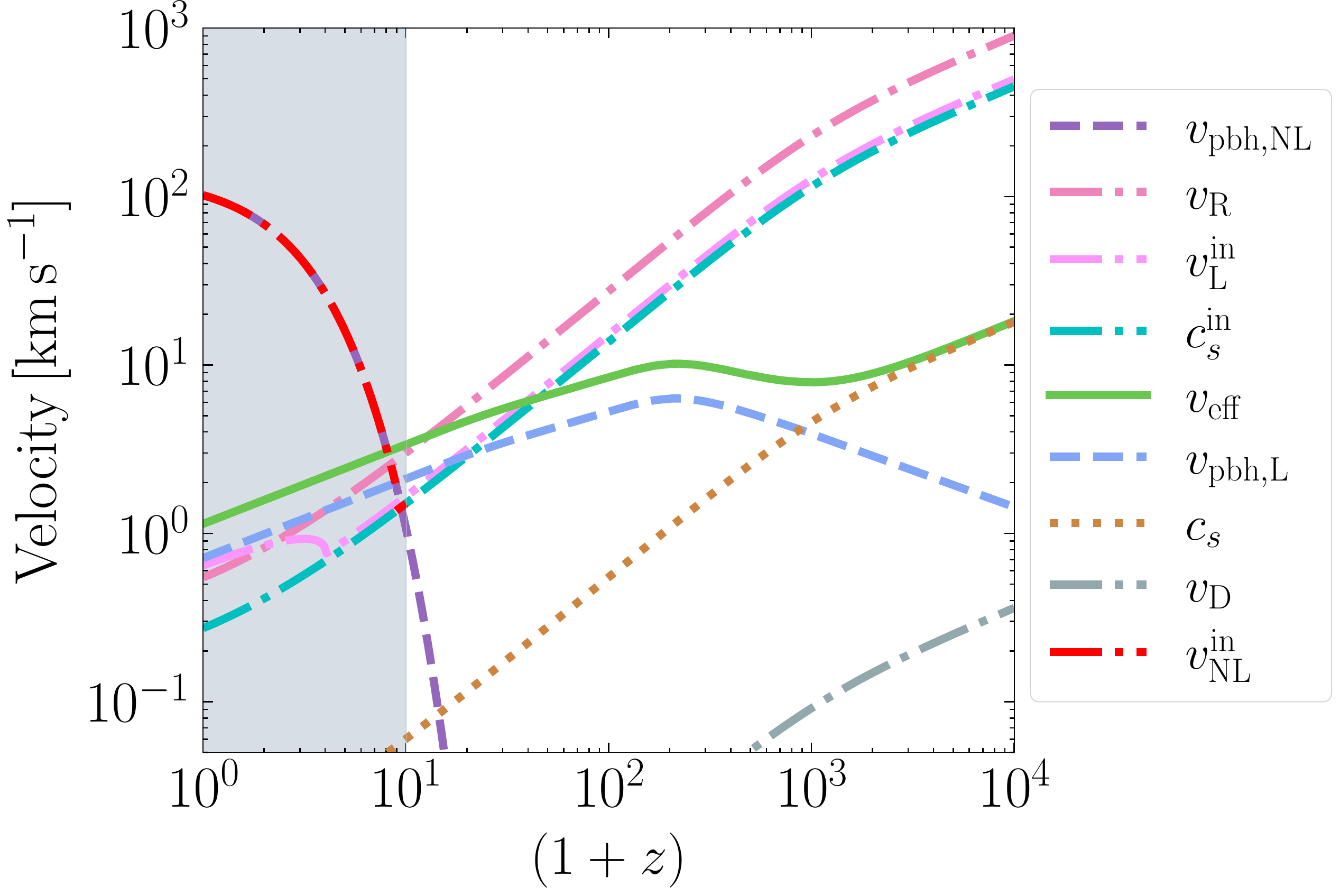}
\caption{Variation of the speed of sound $c_s$, velocity $v_\mathrm{pbh}$ of the PBHs and their effective velocities chosen as per ROM07~\cite{Ricotti:2007au}. Dash-dotted lines correspond to the velocity profiles inside the ionized region in the PR accretion model and the remaining lines show the corresponding values in the neutral region in PR model as well as in the BHL model. Here, the velocity of the PBHs is approximated as the the relative velocity of the baryonic gas and dark matter particles in the Universe.}
\label{fig:velROM07BHLPR}
\end{figure}

In Figure~\ref{fig:velROM07BHLPR}, we show the variation of the velocity profiles (speed of sound $c_s$, PBHs velocity $v_\mathrm{pbh}$ and effective velocities) in the BHL and PR models, based on the expressions above from ROM07. In contrast to the velocity profiles of SPIK20 (see Fig.~\ref{fig:velSPIK20}), the PBH velocity in the linear regime $v_\mathrm{pbh,L}$ (shown by the blue dashed line) first increases up to the redshift of decoupling and then decreases with decrease in $z$. This is due to the fact that the velocity of baryons becomes relatively higher for $z < z_\mathrm{dec} \approx 1100$ which reduces the value of $v_\mathrm{pbh} \equiv \left\langle  v_\mathrm{DM} \right\rangle - \left\langle  v_\mathrm{gas} \right\rangle$ post-decoupling. This typically leads to larger values of the effective velocity in the BHL model (solid green) compared to SPIK20, especially at intermediate redshift. 
Moreover, inside the ionized region in the PR model, we find that  $v_\mathrm{D} < v_\mathrm{pbh} < v_\mathrm{R}$, resulting in a gas density inside the ionized region of $\rho^\mathrm{in} = \rho^\mathrm{in}_{0}$, as for SPIK20. However, below $z \lesssim 5$ in the linear regime, the PBH velocity exceeds $v_\mathrm{R}$, resulting in a change in the gas density $\rho^\mathrm{in} = \rho^\mathrm{in}_{-}$ and an increase in the PBH velocity inside the ionized bubble $v^\mathrm{in}_\mathrm{L}$. The change is even more pronounced in the non-linear regime.

\begin{figure}[tb!]
\centering
\includegraphics[width = 0.48\textwidth]{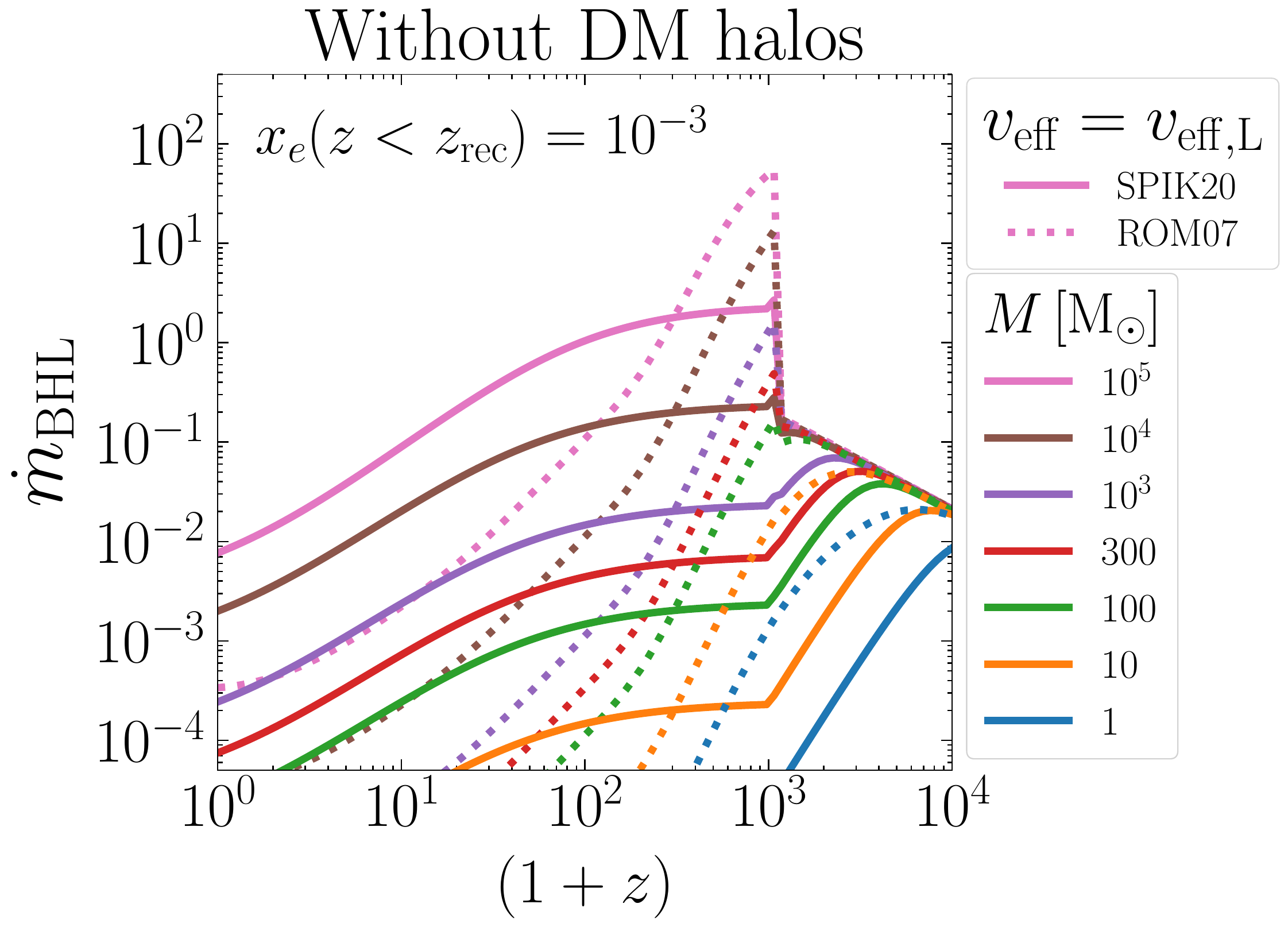}
\hspace*{\fill}
\includegraphics[width = 0.48\textwidth]{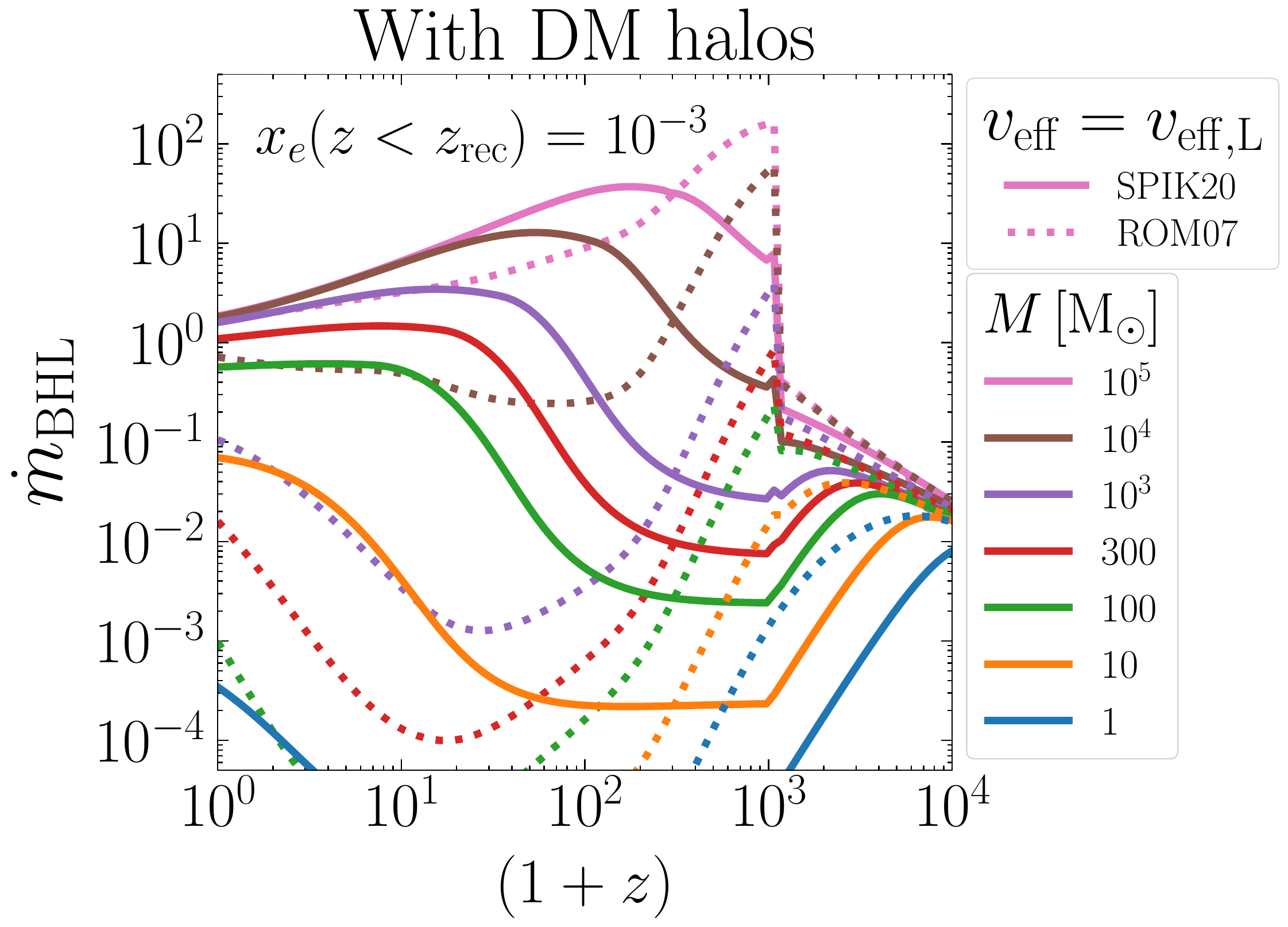}
\caption{Variation of the dimensionless baryonic accretion rates of PBHs with and without DM halos, in BHL model, as a function of redshift $(1+z)$. The dotted (solid) lines represent the accretion rates based on the velocity profiles selected from ROM07~\cite{Ricotti:2007au} (SPIK20~\cite{Serpico:2020ehh}). The left panel depicts the accretion scenarios for isolated PBHs while the right panel shows the corresponding values in the presence of DM halos.}
\label{fig:mdotcompBHL}
\end{figure}

\begin{figure}[tb!]
\centering
\includegraphics[width = 0.48\textwidth]{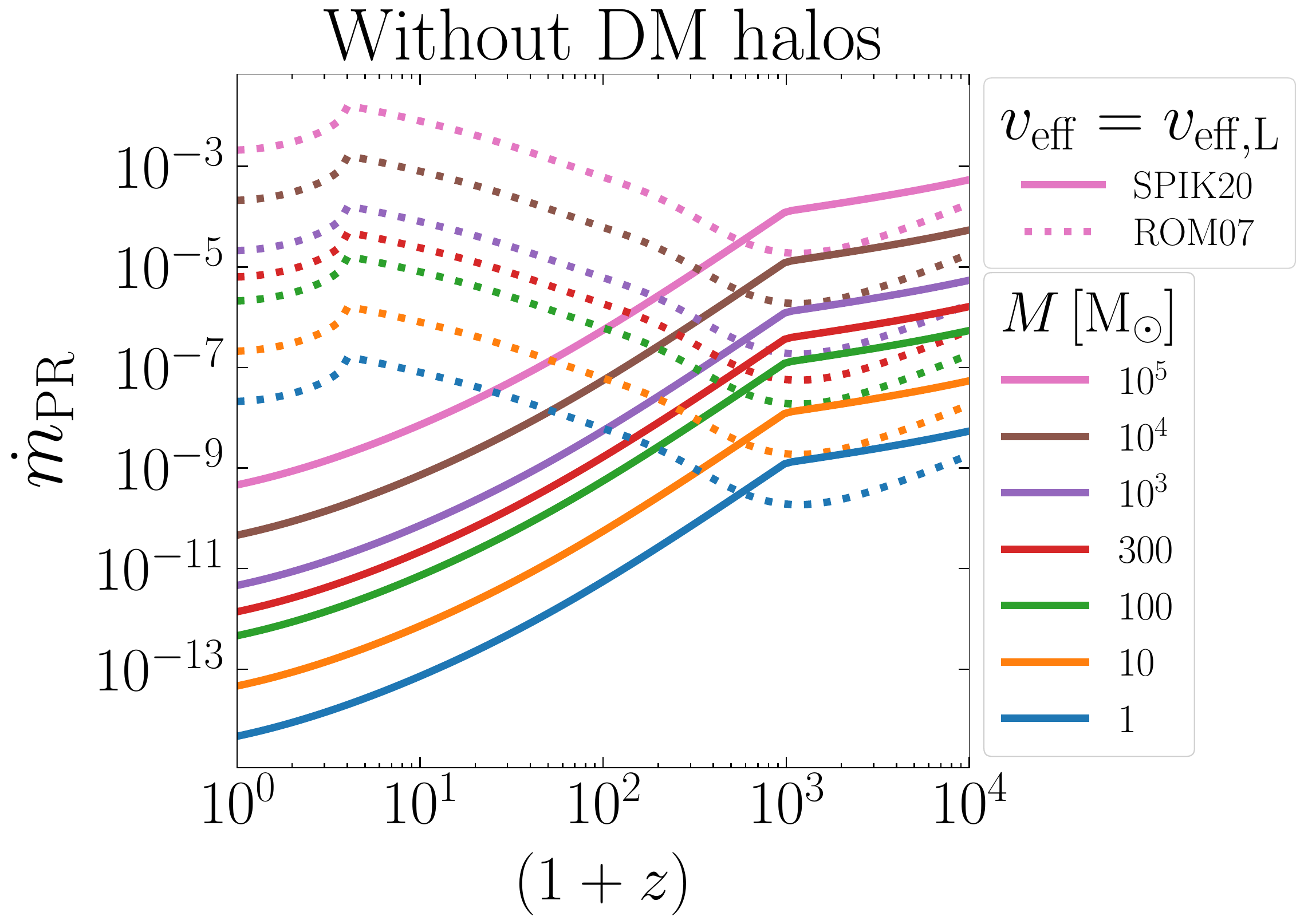}
\vspace*{\fill}
\includegraphics[width = 0.48\textwidth]{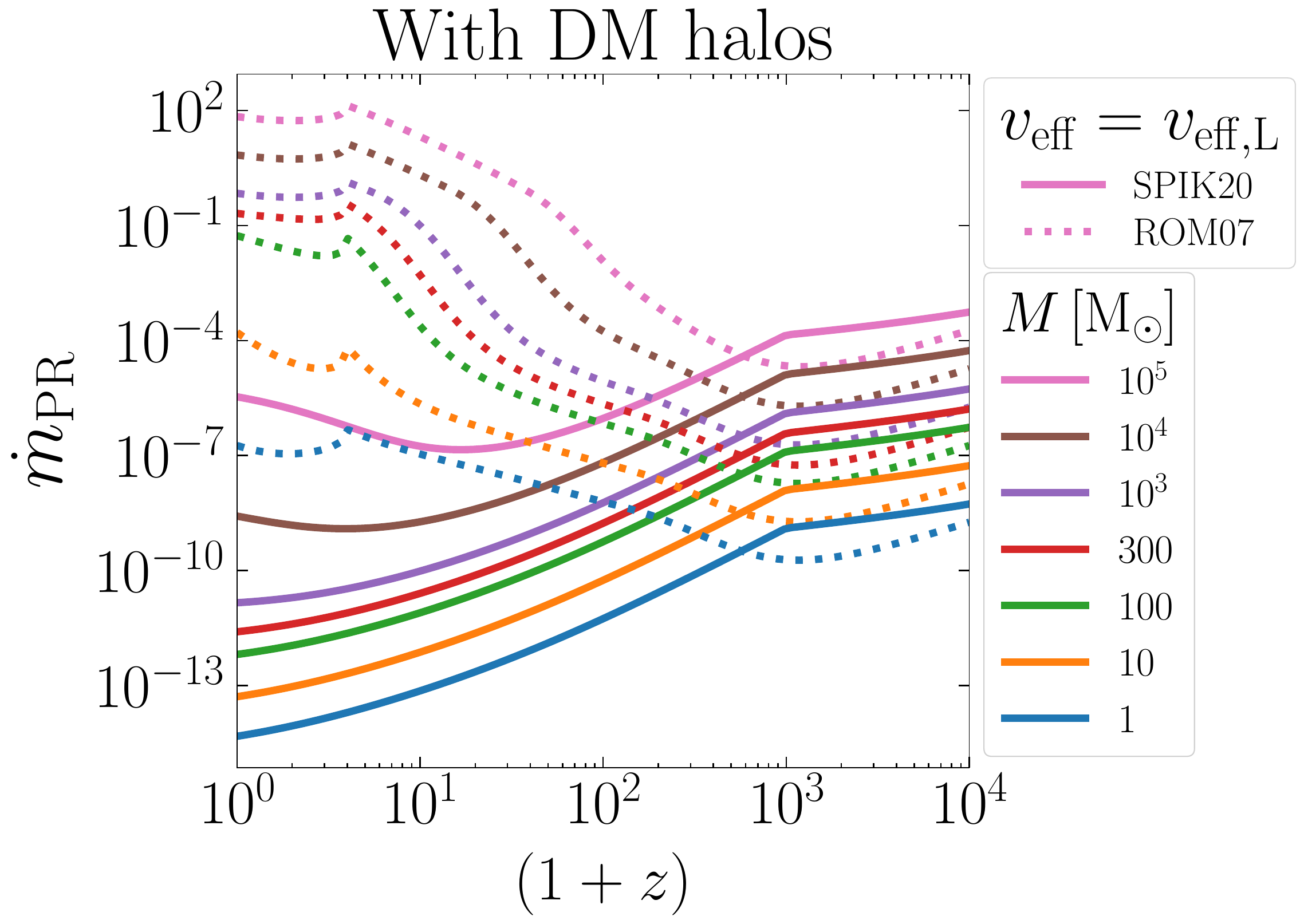}
\caption{Variation of the dimensionless accretion rates of PBHs with and without DM halos within the framework of PR accretion model. The dotted (solid) lines signify the accretion scenario involving the speed of sound $c_s$ and velocity $v_\mathrm{pbh}$ of PBHs selected from ROM07~\cite{Ricotti:2007au} (SPIK20~\cite{Serpico:2020ehh}). The left panel depicts the accretion rates of PBHs without DM halos while the right panel illustrates their counterpart values when the PBHs are surrounded by DM halos.}
\label{fig:mdotcompPR}
\end{figure}

We now compare the accretion rates of PBHs with and without DM halos in the BHL model detailed in Section~\ref{sec:Bondi-Hoyle-Lyttleton (BHL) Accretion model for baryons}, employing the velocity profiles selected as per SPIK20~\cite{Serpico:2020ehh} and ROM07~\cite{Ricotti:2007au}.
For constant electron fraction $x_e = 10^{-3}$ for $z<z_\mathrm{rec}$, the variation of the dimensionless accretion rate $\dot{m}$ of PBHs is shown in Figure~\ref{fig:mdotcompBHL}. The left panel of this figure illustrates that in the linear regime, at high redshift, the accretion rates of isolated PBHs (i.e.\ PBHs without DM halos) are roughly one order of magnitude higher when the velocity profiles are chosen in accordance with ROM07. This is due to the smaller value of $v_\mathrm{eff}$ and therefore larger value of the Bondi radius, compared to SPIK20. However, for low redshift $z \ll z_\mathrm{rec}$, the effective PBH velocity given by ROM07 drops more slowly than in SPIK20, leading to a strong suppression of the accretion rate in the former case. The right panel of Fig.~\ref{fig:mdotcompBHL} shows similar behaviour, though as expected, in both cases the dimensionless accretion rates are larger with the inclusion of DM halos at low redshift. For the largest PBH masses, the accretion rates for SPIK20 and ROM07 are much closer when DM halos are included. This occurs because for heavier PBH masses, the accretion efficiency is not saturated at $\lambda = 1$ (see Fig.~\ref{fig:lambda}) and the higher effective velocity in the ROM07 case leads to a mild increase in the accretion efficiency, bringing the two velocities profiles closer together.\footnote{We note that the accretion rates we obtain here are somewhat smaller than those reported in ROM07 and in Ref.~\cite{DeLuca:2020fpg} (which follows the same formalism as ROM07). The origin of this discrepancy is not clear, but may arise from the choice of $v_\mathrm{eff}$; the accretion rates differ depending on whether an average over $f_M(v, \sigma)$ is taken.}

Figure~\ref{fig:mdotcompPR} compares the dimensionless accretion rates in the PR model assuming velocity profiles from ROM07 and SPIK20. The most notable feature is that at low redshift the ROM07 accretion rates exceed those from SPIK20 by up to 5 orders of magnitude. This is due to the fact that the background sound speed in the ROM07 model drops more rapidly than assumed in SPIK20. This in turns leads to a smaller sound speed and a larger density inside the ionized bubble. Below $z \lesssim 5$, we see a sudden drop in the accretion rate. As described above, this marks the transition to the regime where $v_\mathrm{pbh} > v_\mathrm{R}$. Physically this corresponds to an increase in gas pressure resulting from the bow shock at the ionization front (see Figure~\ref{fig:PRmodel}), leading to a drop in density inside the ionized region.

In the main text, in Figure~\ref{fig:MfBHLPRROM07}, we also show a comparison of the accreted mass of PBHs assuming either ROM07 or SPIK20 velocity profiles. We find that the choice of velocity profile may alter the accreted mass by up to 5 orders of magnitude at low redshift. This highlights the fact that the speed of sound and the velocity of the PBHs critically influence the efficiency of accretion around PBHs and their subsequent evolution in the Universe. The accurate modeling of these parameters is very important to fully comprehend the theoretical frameworks of accretion models for PBHs and their potential observational signatures.

\clearpage
\bibliographystyle{JHEP}
\bibliography{acc}
\end{document}